%% file: agpd.tex
\documentclass{article}

\PassOptionsToPackage{numbers, compress}{natbib}

 \usepackage[preprint]{neurips_2024}

\usepackage[utf8]{inputenc} 
\usepackage[T1]{fontenc}    
\usepackage{hyperref}       
\usepackage{url}            
\usepackage{booktabs}       
\usepackage{amsfonts}       
\usepackage{nicefrac}       
\usepackage{microtype}      
\usepackage{xcolor}         
\usepackage{xspace}
\usepackage{multirow}
\usepackage{sidecap}
\usepackage{amsmath}
\usepackage{bm}
\usepackage{array}
\usepackage{booktabs}
\usepackage{algorithm}
\usepackage{algorithmic}
\newtheorem{definition}{Definition}
\definecolor{snsblue}{rgb}{0.198, 0.347, 0.590}
\definecolor{darkblue}{RGB}{0,0,180} 
\definecolor{darkred}{RGB}{180,0,0} 
\definecolor{darkpurple}{RGB}{120,0,180} 
\definecolor{darkgreen}{RGB}{0,120,0} 
\definecolor{darkbrown}{RGB}{100,40,0} 
\definecolor{grey}{RGB}{128,128,128} 

\usepackage[capitalize]{cleveref}
\crefname{section}{Sec.}{Secs.}
\Crefname{section}{Section}{Sections}
\Crefname{table}{Table}{Tables}
\crefname{table}{Tab.}{Tabs.}
\usepackage{graphicx}
\usepackage{caption}
\usepackage{wrapfig}
\usepackage{subcaption}
\usepackage{colortbl}
\usepackage{enumitem}
\newcommand{\ie}{\textit{i.e.,}\xspace}
\newcommand{\eg}{\textit{e.g.,}\xspace}
\newcommand{\wrt}{\textit{w.r.t.}\xspace}

\newcommand{\x}{\bm{x}} 
\newcommand{\h}{\bm{h}} 
\newcommand{\g}{\bm{g}} 
\newcommand{\D}{\mathcal{D}}
\newcommand{\X}{\mathcal{X}}
\newcommand{\Y}{\mathcal{Y}}
\newcommand{\etc}{\textit{etc}}
\title{Activation Gradient based Poisoned Sample Detection Against Backdoor Attacks}

\author{Danni Yuan\textsuperscript{1}, Shaokui Wei\textsuperscript{1}, Mingda Zhang\textsuperscript{1}, Li Liu\textsuperscript{2}, Baoyuan Wu\textsuperscript{1\thanks{Corresponds to Baoyuan Wu \href{mailto:wubaoyuan@cuhk.edu.cn}{(wubaoyuan@cuhk.edu.cn)}}}\\
\textsuperscript{1}School of Data Science,
The Chinese University of Hong Kong, Shenzhen, \\Guangdong, 518172, P.R. China\\
\textsuperscript{2}The Hong Kong University of Science and Technology (Guangzhou),\\ Guangzhou, China
}

\begin{document}

\maketitle

\input{section/abstract}
\input{section/introduction}
\input{section/related_work}
\input{section/gcd}

\input{section/detection-method}
\input{section/experiment}
\input{section/conclusion}
\bibliographystyle{plainnat}
\bibliography{main}
\newpage
\input{section/appendix}
\end{document}

%% file: section/abstract.tex
\begin{abstract}

This work studies the task of poisoned sample detection for defending against data poisoning based backdoor attacks. Its core challenge is finding a generalizable and discriminative metric to distinguish between clean and various types of poisoned samples (\eg various triggers, various poisoning ratios). Inspired by a common phenomenon in backdoor attacks that the backdoored model tend to map significantly different poisoned and clean samples within the target class to similar activation areas, we introduce a novel perspective of the circular distribution of the gradients \wrt sample activation, dubbed \textit{gradient circular distribution} (GCD). And, we find two interesting observations based on GCD. One is that the GCD of samples in the target class is much more dispersed than that in the clean class. The other is that in the GCD of target class, poisoned and clean samples are clearly separated. Inspired by above two observations, we develop an innovative three-stage poisoned sample detection approach, called \textit{Activation Gradient based Poisoned sample Detection (AGPD)}. First, we calculate GCDs of all classes from the model trained on the untrustworthy dataset. Then, we identify the target class(es) based on the difference on GCD dispersion between target and clean classes. Last, we filter out poisoned samples within the identified target class(es) based on the clear separation between poisoned and clean samples. Extensive experiments under various settings of backdoor attacks demonstrate the superior detection performance of the proposed method to existing poisoned detection approaches according to sample activation-based metrics. 
\end{abstract}

%% file: section/introduction.tex
\section{Introduction}
\label{sec: intro}

It is well known that deep neural networks (DNNs) are vulnerable to backdoor attacks \cite{wu2023adversarial}, where the adversary could inject a particular backdoor into the DNN model through manipulating the training dataset or training process. 
Consequently,the backdoored model will produce a target label when encountering a particular trigger pattern, leading to unexpected security threats in practice. 
Protecting DNNs from backdoor attacks is an urgent and important task. 

Here we focus on defending against the data-poisoning based backdoor attacks by filtering out the potential poisoned samples from a untrustworthy training dataset, \ie \textit{poisoned sample detection (PSD)}. One of the main challenges for PSD is the information lack of the potential poisoned samples, such as the trigger type, the target class(es), the number of poisoned samples, \etc. Some seminal works have been developed by exploring some discriminative metrics based on the intermediate activation or predictions of poisoned and clean samples in the backdoored model trained on the untrustworthy dataset, such as activation clustering (AC)~\cite{ac}, STRIP~\cite{gao2019strip}, SCAn~\cite{tang2021demon}. However, the recently proposed adaptive backdoor poisoning attack~\cite{adapt} challenges the assumption that poisoned and clean samples can be distinctly separated in activation space. 

\begin{wrapfigure}{r}{0.5\textwidth}
  \vspace{-0.25in}
  \begin{center}
    \includegraphics[width=0.5\textwidth]{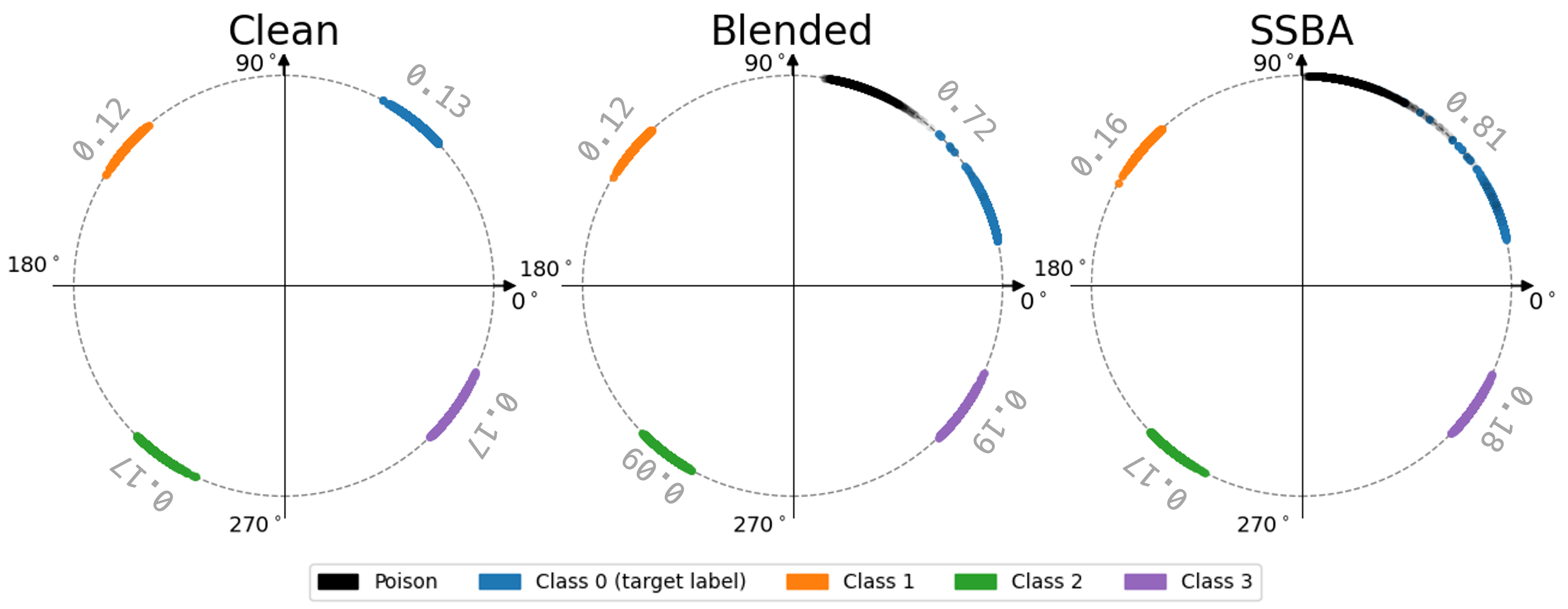}
  \end{center}
  \vspace{-0.15in}
   \caption{Gradient circular distributions (GCDs) across four classes of CIFAR-10, on the clean model \textbf{(left)}, Blended attacked model \textbf{(middle)}, and SSBA attacked model \textbf{(right)}, respectively. The value along with each arc indicates the CVBT value. The GCD of the target class (covering both \textbf{black} and \textbf{\textcolor{snsblue}{blue}} arcs). Note that we moved three clean classes' arcs to different quadrants to avoid visual overlap.}
   \label{fig: gcd}
  \vspace{-0.1in}
\end{wrapfigure}

In this work, we introduce a novel perspective that distinguishes the behavior of poisoned and clean samples by tracking their activation gradients (\ie the gradient \wrt activation). It is inspired by the phenomenon that a backdoored model tends to map both poisoned and clean samples within the target class to similar areas in its activation space \cite{huang2022backdoor}, such that they can be predicted as the same label. Considering the significant discrepancy between poisoned and clean samples in their original input space, \textit{their mapping directions should be significantly different}, while the mapping direction could be reflected by the activation gradient direction. 
Thus, we define a new concept called \textit{gradient circular distribution (GCD)} (introduced in Section \ref{sec: preliminary gcd}), to capture the distribution of activation gradient directions. 
Take Fig. \ref{fig: gcd} as the example, given a trained model, we calculate one GCD of training samples in each class. 
There are \textbf{two interesting observations}: 
\vspace{-.5em}
\begin{itemize}[leftmargin=1em]
    \item \textbf{Observation 1 on GCD dispersion}:  Given one backdoored model (see the middle/right sub-figure), the GCD of the target class is much more dispersed than GCDs of all clean classes;
    \vspace{-.3em}
    \item \textbf{Observation 2 on sample separation in target GCD}: In the GCD of target class, poisoned and clean samples are clearly separated (see the \textbf{black} and \textbf{\textcolor{snsblue}{blue}} arcs in the middle/right sub-figure).
    \vspace{-.5em}
\end{itemize}

Motivated by above two observations, we develop an innovative poisoned sample detection approach, called \textbf{Activation Gradient based Poisoned sample Detection (AGPD)}, which consist of three stages.  
\textbf{First}, we train a DNN model based on the untrustworthy dataset, and calculate GCDs of all classes. 
\textbf{Second}, we identify the target class(es) according to a novel class-level metric that measures the dispersion of each class's GCD (corresponding to the first observation). 
\textbf{Last}, within the identified target class(es), we gradually filter out poisoned samples according to a novel sample-level metric that measures the closeness to the clean reference sample (corresponding to the second observation). 

Moreover, we conduct extensive evaluations under various backdoor attacks and various datasets, and show that the activation gradient is more discriminative than the activation to distinguish between poisoned and clean samples, which explains the superior

In summary, the \textbf{main contributions} of this work are three-fold. 
\textbf{(1)} We introduce a novel perspective for poisoned sample detection, called gradient circular distribution (GCD), and present two interesting observations based on GCD. 
\textbf{(2)} We develop an innovative approach by sequentially identifying the target class(es) and filtering poisoned samples for the poisoned sample detection task, based on GCD and two novel metrics about GCD. 
\textbf{(3)} We conduct extensive evaluations and analysis to verify the superiority of the proposed approach to existing activation-based detection approaches. 

%% file: section/related_work.tex
\section{Related work}
\paragraph{Backdoor attack.} BadNets~\cite{gu2019badnets} is the pioneering work that introduces the concept of backdoor attack into Deep Neural Networks (DNNs), in which the adversary manipulates training samples by adding a small patch with specific patterns and changing their labels to a target label. 
Following this, the variety of triggers expanded significantly, including a cartoon image used in Blended~\cite{chen2017targeted}, a universal adversarial perturbation with only low-frequency components utilized in Low-Frequency~\cite{zeng2021rethinking}, and a sinusoidal signal employed in SIG~\cite{SIG}, \etc. 
One commonality of these triggers is that they are sample-agnostic, \ie they are same across different poisoned samples. This characteristic is often exploited to effectively detect and remove the triggers or the poisoned samples. 
To counteract this, sample-specific triggers have been designed, such as WaNet~\cite{nguyen2021wanet}, Input-Aware~\cite{nguyen2020input}, SSBA~\cite{li2021invisible}, CTRL~\cite{li2023embarrassingly}, TaCT~\cite{tang2021demon}, and Adap-Blend~\cite{adapt}. 
These attacks use more complex and dynamic triggers, posing significant challenges for poisoned sample detection. 
Additionally, some attacks explore various attack settings regarding the number of triggers and target classes, such as all-to-all attack (\eg BadNets-A2A~\cite{gu2019badnets}), multi-target and multi-trigger attack (\eg c-BaN~\cite{salem2022dynamic}). These diverse settings further complicate the detection of poisoned samples.

\paragraph{Backdoor defense.} 
According to the accessible information, several different branches of backdoor defense methods have been developed, such as the pre-training backdoor defense (\eg \cite{ac,tran2018spectral,al2023don}) if given a untrustworthy training dataset, in-training backdoor defense (\eg \cite{huang2022backdoor,li2021anti,chen2022effective, mu2023progressive, gao2023effectiveness}) if the training process can be controlled by defender, as well as post-training backdoor defense (\eg \cite{liu2018fine,zhu2023neural,zhu2023enhancing,wei2023shared,wang2019neural,wu2021adversarial,zeng2022adversarial,zheng2022preactivation, chai2022oneshot,zheng2022data}) if given a backdoored model. Due to space limitations, we will only review existing methods of poisoned sample detection (PSD), which belong to pre-training backdoor defense. Most existing PSD methods aim to construct discriminative metrics between poisoned and clean samples based on intermediate activations, final predictions, or loss values. For activation-based methods, such as activation clustering (AC) \cite{ac}, Beatrix \cite{mabeatrix}, SCAn \cite{tang2021demon}, and Spectral \cite{tran2018spectral}, they utilize dimensionality reduction and clustering techniques, such as K-means clustering, Gram matrix analysis, two-component decomposition, and SVD, to distinguish poisoned and clean samples. For input-based methods, STRIP \cite{gao2019strip} uses the entropy of predictions on perturbed inputs to identify poisoned samples, while CD \cite{huang2023distilling} measures the $L_1$ norm of the learned masks on inputs to detect poisoned samples. For loss-based methods, like ABL \cite{li2021anti} and ASSET \cite{pan2023asset}, they observed that the loss of poisoned samples decreases quickly during the early training epochs, leveraging this phenomenon to identify poisoned samples.

%% file: section/gcd.tex
\section{Preliminary: gradient circular distribution}
\label{sec: preliminary gcd}

\paragraph{Task setting.} In this section, given a DNN-based classification model $f_{\boldsymbol{w}}: \mathcal{X} \rightarrow \mathcal{Y}$, with $\X \in \mathbb{R}^d $ being the input sample space and $\Y = \{1, 2, \ldots, K\}$ being the output space with $K$ candidate classes, as well as a dataset $\D = \{(\x_i, y_i)\}_{i=1}^n$, we investigate their gradients in the model $f_{\boldsymbol{w}}$. 

\subsection{Definition of gradient circular distribution}

Here we introduce the definitions of Activation Gradient and Gradient Circular Distribution (GCD), as described in Definition \ref{def:ag} and Definition \ref{def:gcd}, respectively. 

\begin{definition}[Activiation Gradient] 
Given a model $f_{\boldsymbol{w}}$, for a sample $\x$ labelled as $y$, we denote its activation map at the $l$-th layer as $\h^{(l)}_{\x}\in\mathbb{R}^{C^{(l)}\times H^{(l)}\times W^{(l)}}$, where $C^{(l)}$, $H^{(l)}$, $W^{(l)}$ are its depth (number of channels), height and width, respectively. Then, we define the \textbf{channel-wise activation gradient}
$\g^{(l)}(\x, y)\in\mathbb{R}^{C^{(l)}}$ as
\begin{equation}
	\g_{\boldsymbol{w}}^{(l)}(\x, y) = \frac{1}{H^{(l)}W^{(l)}}\sum_{h=1}^{H^{(l)}}\sum_{w=1}^{W^{(l)}}\frac{\partial [f_{\boldsymbol{w}}(\x)]_y}{\partial{[\h^{(l)}_{\x}]_{:,h,w}}} \in\mathbb{R}^{C^{(l)}},
 \label{eq: activation gradient}
\end{equation}
where $[f_{\boldsymbol{w}}(\x)]_y$ is the logit \wrt class $y$ and $[\h^{(l)}_{\x}]_{:,h,w}\in\mathbb{R}^{C^{(l)}}$ is the activation sliced at height $h$ and width $w$ over all channels. 
For simplicity, if no special specifications are required, hereafter we will refer to it as $\g(\x)$.
\label{def:ag}
\end{definition}

\begin{definition}[Gradient Circular Distribution (GCD)] 
Given a model $f_{\boldsymbol{w}}(\cdot)$, a set of samples $\mathcal{D} = \{(\x_i, y_i) \}_{i=1}^n$, and a basis sample pair $(\x_0, y_0)$, we firstly calculate the activation gradient of each sample, \ie $\g(\x_i), i=0,1,\ldots,n$. 
 
Then, take $\g_{\x_0}$ as the (unnormalized) basis vector, the angle of each sample in $\mathcal{D}$ is calculated as follows: 
\begin{equation}
\theta_{\x_0}(\x_i)
= \arccos \left( \frac{\g(\x_i) \cdot \g(\x_0)}{\|\g(\x_i)\| \|\g(\x_0)\|} \right) \in [0, 2\pi), i=1,\ldots,n, 
\label{eq: gcd}
\end{equation}
where $\cdot$ denotes the dot product, and $\|\cdot \|$ returns the magnitude. 

The distribution of the angle set $\{\theta_{\x_0}(\x_i)\}_{i=1}^n$ is called as the gradient circular distribution (GCD) of the set $\mathcal{D}$, denoted as 
$\mathcal{P}_{\x_0}(\mathcal{D})$.

\label{def:gcd}
\end{definition}

\subsection{Characteristics of gradient circular distribution}

Furthermore, to quantize the characteristics of $\mathcal{P}_{\x_0}(\mathcal{D})$, we introduce the following two metrics. 

\paragraph{Dispersion metric of $\mathcal{P}_{\x_0}(\mathcal{D})$.} 
To measure the dispersion of $\mathcal{P}_{\x_0}(\mathcal{D})$, we design a novel metric called 
\textit{\textbf{\underline{C}}osine similarity \textbf{\underline{V}}ariation towards \textbf{\underline{B}}asis \textbf{\underline{T}}ransition} (\textbf{CVBT}). 
Specifically, given $\{\theta_{\x_0}(\x_i)\}_{i=1}^n$, we firstly pick the activation gradient $\g_{\x_{n^*}}$ corresponding to the largest angle, \ie $n^* = \arg\max_{i\in \{1,\ldots,n\}} \theta_{\x_0}(\x_i)$. 
In other words, $\g(\x_{n^*})$ is the farthest activation gradient vector from the original basis vector $\g(\x_{0})$. 
Then, by setting $\g(\x_{n^*})$ as a new basis vector, we calculate $\{\theta_{\x_{n^*}}(\x_i)\}_{i=1}^n$ using Eq. (\ref{eq: gcd}). Based on $\{\theta_{\x_0}(\x_i)\}_{i=1}^n$ and $\{\theta_{\x_{n^*}}(\x_i)\}_{i=1}^n$, we formulate the CVBT metric of $\mathcal{P}_{\x_0}(\mathcal{D})$ as follows:
\begin{equation}
 \rho_{\x_{0}}(\mathcal{D}) = \bigg(\frac{1}{n} \sum_{i=1}^n \big(\cos(\theta_{\x_{0}}(\x_i))-\cos(\theta_{\x_{n^*}}(\x_i)) \big)^2 \bigg)^{\frac{1}{2}} \in [0,2], 
    \label{eq: cvbt}
\end{equation}
where $\cos(\theta)$ returns the cosine value of an angle $\theta$. 
Note that \textbf{larger $\rho_{\x_{0}}(\mathcal{D})$ indicates larger dispersion of $\mathcal{P}_{\x_0}(\mathcal{D})$}. 
For example, 
if the farthest vector $\g(\x_{n^*})$ has the same direction with $\g(\x_{0})$, implying $\{\theta_{\x_0}(\x_i)\}_{i=1}^n$ are all zeros, then $\rho_{\x_{0}}(\mathcal{D})=0$, meaning the zero dispersion of $\mathcal{P}_{\x_0}(\mathcal{D})$. 
In contrast, if $\g(\x_{n^*})$ has the opposite direction with $\g(\x_{0})$, and all remaining $\g(\x_{i})$ are either same or opposite directions with $\g(\x_{0})$, then $\rho_{\x_{0}}(\mathcal{D})=2$, meaning the largest dispersion of $\mathcal{P}_{\x_0}(\mathcal{D})$.

\paragraph{Sample-level closeness metric based on $\mathcal{P}_{\x_0}(\mathcal{D})$.}  
Given $\{\theta_{\x_0}(\x_i)\}_{i=1}^n$ and $\{\theta_{\x_{n^*}}(\x_i)\}_{i=1}^n$, we design a novel metric to measure the closeness of each sample $\x_i$ to the reference sample $\x_0$, as follows: 
\begin{equation}
\label{eq:trust_score}
s_{\x_0}(\x_i) = \frac{1-\cos(\theta_{\x_{n^*}}(\x_i))}{(1-\cos(\theta_{\x_{n^*}}(\x_i)) + (1-\cos(\theta_{\x_{0}}(\x_i))} \in [0, 1).
\end{equation}
Note that \textbf{larger $s_{\x_0}(\x_i)$ indicates greater closeness of $\x_i$ to $\x_0$}. For example, if $\g(\x_{i})$ has the same direction with $\g(\x_{n^*})$ while the opposite direction with $\g(\x_{0})$, then $s_{\x_0}(\x_i) =0$, implying the farthest from $\x_0$. 
In contrast, if $\g(\x_{i})$ has the opposite direction with $\g(\x_{n^*})$ while the same direction with $\g(\x_{0})$, then $s_{\x_0}(\x_i)=1$, implying the closest to $\x_0$.

%% file: section/detection-method.tex
\section{Activation gradient based poisoned detection method}

\subsection{Problem setting}

\paragraph{Threat model.} We consider the threat model of data poisoning based backdoor attack. The adversary generates a poisoned dataset $\D_{bd}$, containing a clean subset $\mathcal{D}_{c} = \{(\x_i, y_i)\}_{i=1}^{n_c}$ and a poisoned subset $\mathcal{D}_{p} = \{(\tilde{\x}_i, t)\}_{i=1}^{n_p}$. $\x, \tilde{\x} \in \mathcal{X}$ denotes the clean and poisoned sample with trigger, respectively. $y, t \in \mathcal{Y}$ indicates the ground-truth and target label, respectively. 
We denote $r = \frac{n_p}{n_c + n_p}$ as the poisoning ratio. 
Note that there could be multiple triggers (\ie multi-trigger) and multiple target labels (\ie multi-target) in the poisoned subset.

\paragraph{Defender's goal.} The defender aims to identify poisoned samples from the untrustworthy dataset $\mathcal{D}_{bd}$. 
We assume that the defender has access to $\mathcal{D}_{bd}$, and a small set of additional clean samples $\mathcal{D}_{ac}$, which contains at least one clean sample for each class. 
Besides, the defender has the capability to train a DNN model $f_{\boldsymbol{w_{bd}}}: \mathcal{X} \to \Y$ based on $\mathcal{D}_{bd}$.

\begin{figure}[h]
  \centering
\includegraphics[width=1\linewidth]{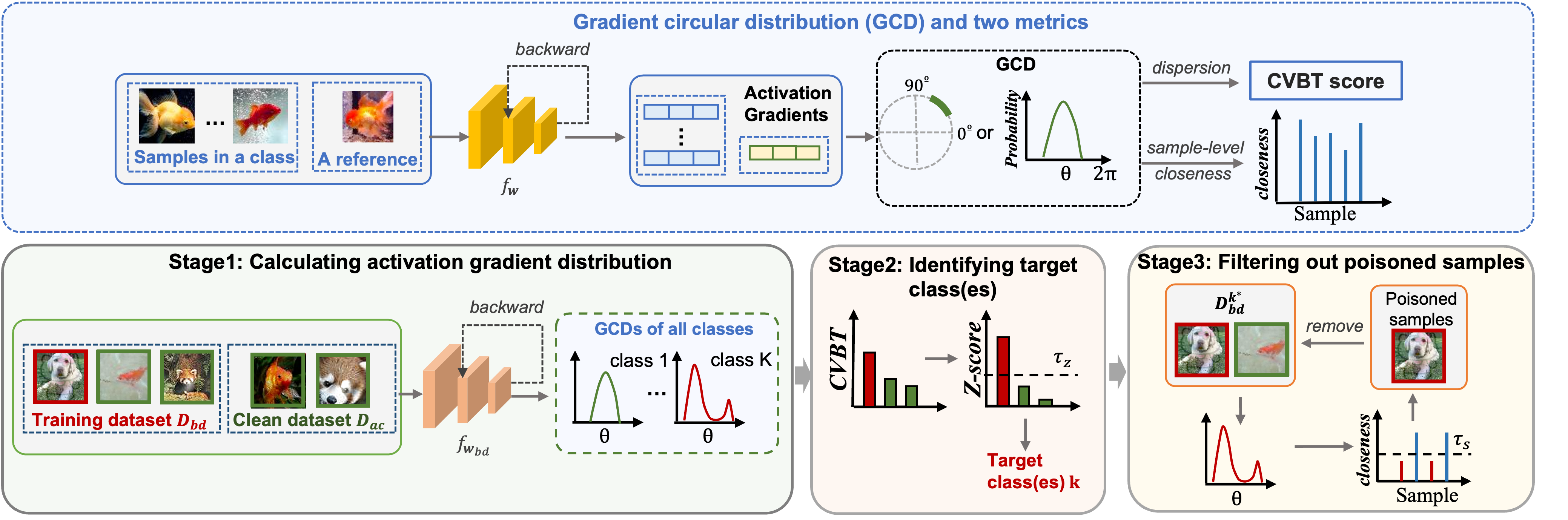}
   \caption{Illustrations of gradient circular distribution (GCD) and two metrics, and the pipeline of the proposed APGD method which consists of three stages: 1) calculating activation gradient distribution, 2) identifying target class(es), and 3) filtering out poisoned samples within the identified target class(es).}
   \label{fig: framework}
\end{figure}

\subsection{Poisoned sample detection method}

Inspired by the two observations demonstrated in Section \ref{sec: intro} and Fig. \ref{fig: gcd}, we develop an innovative poisoned sample detection method by utilizing GCD and the corresponding metrics (see Section \ref{sec: preliminary gcd}), called \textit{\underline{A}ctivation \underline{G}radient based \underline{P}oisoned \underline{D}etection} (\textbf{AGPD}).
As illustrated in Fig. \ref{fig: framework}, AGPD consists of three stages, as detailed below.

\paragraph{Stage 1: Calculating activation gradient distribution.} 
We denote the samples of class $k$ in $\D_{bd}$ as $\D_{bd}^{k} = \{ (\x_i, k)\}_{i=1}^{n_k}$. 
Given the model $f_{\boldsymbol{w_{bd}}}$ trained on $\D_{bd}$ (the training details will be provided in Appendix \ref{appendix:hyper}), and picking one clean sample pair $(\x_0^k, k) \in \D_{ac}$ as the reference, we can calculate the GCD of $\D_{bd}^{k}$, according to Eqs. (\ref{eq: activation gradient}) and (\ref{eq: gcd}). Consequently, we obtain $\{\mathcal{P}_{\x_0^k}^{(l)}(\D_{bd}^{k})\}_ {l, k=1, 1}^{L, K}$. 
Note that as defined in Eq. (\ref{eq: activation gradient}), the superscript $(l)$ indicates that we adopt the activation gradients of the $l$-th layer in $f_{\boldsymbol{w_{bd}}}$ to calculate GCD. 
For simplicity, hereafter we denote $\mathcal{P}_{\x_0^k}^{(l)}(\D_{bd}^{k})$ as $\mathcal{P}_{k}^{(l)}$.

\paragraph{Stage 2: Identifying target class(es).} 
According to the aforementioned first observation, the GCD of the target class is likely to be more dispersed than that of the clean class. 
Thus, we firstly calculate the dispersion value $\rho_{\x_0^k}^{(l)}(\D_{bd}^{k})$ (for simplicity, we denote it as $\rho_k^{(l)}$) of each $\mathcal{P}_{k}^{(l)}$, according to the CVBT metric (see Eq. (\ref{eq: cvbt})). 
As shown in Fig. \ref{fig: framework}, since $\rho_k^{(l)}$ of target class(es) is likely to be larger , while those of clean classes are likely to be small, we can adopt the anomaly detection technique to identify target class(es), such as the absolute robust Z-score \citep{iglewicz1993volume} method. 
Specifically,  
we can calculate the Z-score of $\rho_k^{(l)}$, as follows:
\begin{equation}
\label{eq:mad}
z_k^{(l)} = \frac{\rho_k^{(l)} - \tilde{\rho}^{(l)}}{\gamma \times \operatorname{MAD}(\{\rho_k^{(l)}\}_{k=1}^K)},
\end{equation}
where $\operatorname{MAD}(\{\rho_k^{(l)}\}) = \operatorname{median}(\{|\rho_k^{(l)} - \tilde{\rho}^{(l)}|\})$ indicates 
the median-absolute-deviation (MAD), and $\tilde{\rho}^{(l)}$ denotes the median value of $\{\rho_k^{(l)}\}_{k=1}^K$. $\gamma$ is a statistical constant valued at 1.4826.
Larger $z_k^{(l)}$ indicates larger likelihood of anomaly. 
We firstly choose the layer with the largest Z-score, \ie $l^* = \arg\max_l (\max_k z_k^{(l)})$. Then, \textbf{if $z_k^{(l^*)}$ exceeds some threshold, \ie $z_k^{(l^*)} \geq \tau_z$, $k$ is identified as a target class}, otherwise clean class. $\tau_z$ will be specified in later experiments.

\paragraph{Stage 3: Filtering out poisoned samples within the identified target class(es).} 
Inspired by the second observation mentioned in Section \ref{sec: intro} that the poisoned sample is likely to be far from the clean sample in GCD, here develop a novel algorithm which gradually filters out poisoned samples with the identified target class(es). 
Specifically, as illustrated in Fig. \ref{fig: framework}, when obtained the identified target class $k^*$, we firstly pick one clean sample pair of class $k^*$ from $\D_{ac}$ as the reference $(\x_0, k^*)$, then we conduct the following three steps iteratively, until a stopping criteria is satisfied:
\vspace{-.8em}
\begin{enumerate}[leftmargin=8mm]
    \item For the set $\D_{bd}^{k^*}$, we calculate its GCD according to Definition \ref{def:gcd}, \ie $\mathcal{P}_{k^*}^{(l^*)}$;
    \item We calculate sample-level closeness value $s_{\x_0}(\x_i)$ for each $\x_i \in \D_{bd}^{k^*}$; 
    \item If the closeness value of one sample is lower than the threshold $\tau_s$ (specified in later experiments), \ie $s_{\x_0}(\x_i) < \tau_s$, then this sample is identified as poisoned, as it is far from the clean reference sample $\x_0$. Then,  $\D_{bd}^{k^*}$ is updated by removing these identified poisoned samples.
\end{enumerate}
In terms of the stopping criteria, we propose to firstly conduct the above iterations until $\D_{bd}^{k^*}$ becomes a null set. At each iteration, we calculate the distribution of $\{s_{\x_0}(\x_i)\}_{\x_i \in \D_{bd}^{k^*}}$, and the Jensen–Shannon (JS) divergence between the current and its previous distribution. Then, we adopt a trace-back strategy by checking the JS divergence value of all iterations, and the iteration that its JS divergence locates at the stable and low region could be set as the stopping iteration. 
Due to the space limit, more details of the whole algorithm, as well as the stopping criteria, will be presented in Appendix \ref{appendix:sliding}. 

%% file: section/experiment.tex
\section{Experiments}
\subsection{Experimental setup}

\paragraph{Attack settings.}
To evaluate the performance of our detection method, we conduct 10 state-of-the-art (SOTA) backdoor attacks that cover 4 categories: 1) \textit{non-clean label with sample-agnostic trigger}, such as BadNets~\cite{gu2019badnets}, Blended~\cite{chen2017targeted}, LF~\cite{zeng2021rethinking}; 2) \textit{clean-label with sample-agnostic trigger}, like SIG~\cite{SIG}; 3) \textit{clean-label with sample-specific trigger}, such as CTRL \cite{li2023embarrassingly}, an attack based on self-supervised learning; and 4) \textit{non-clean label with sample-specific trigger}, including SSBA~\cite{li2021invisible},  WaNet~\cite{nguyen2021wanet}, Input-Aware~\cite{nguyen2020input}, TaCT~\cite{tang2021demon}, and Adap-Blend~\cite{adapt}. These attack settings follow BackdoorBench~\cite{wu2022backdoorbench} for a fair comparison. The poisoning ratio in our main evaluation is 10\% for non-clean label attacks and 5\% for clean label attacks. The target label $t$ is set to $0$ for all-to-one backdoor attack, while target labels are set to $t = (y+1) \mod K$ for all-to-all backdoor attack.

\paragraph{Detection settings.} 
We compare AGPD with eight detection methods, categorized into three groups: 1) \textit{activation-based}, including AC~\cite{ac}, Beatrix~\cite{mabeatrix}, SCAn~\cite{tang2021demon}, and Spectral~\cite{tran2018spectral}; 2) \textit{input-based}, such as STRIP~\cite{gao2019strip} and CD~\cite{huang2023distilling}; 3) \textit{loss-based}, represented by ABL~\cite{li2021anti} and ASSET~\cite{pan2023asset}. In this work, we assume that defenders have access to the entire training dataset, while the model's architecture and training process can be independently selected and implemented by the defenders themselves. Furthermore, it is presumed that defenders can gather a small, additional clean dataset to support the detection process, as suggested in previous works~\cite{mabeatrix}\cite{tang2021demon}\cite{gao2019strip}. For a fair comparison, we maintain that the number of clean samples per class is 10, extracted from the test dataset. The threshold used in AGPD $\tau_z$ and $\tau_s$ are $e^2$ and $0.05$, respectively. 

\paragraph{Datasets and models.} 
We use CIFAR-10~\cite{krizhevsky2009learning} and Tiny ImageNet~\cite{le2015tiny} as primary datasets to evaluate the detection performance. Additionally, we expand our evaluation to the datasets that are closer to real-world scenarios, such as ImageNet~\cite{deng2009imagenet} subset (200 classes), DTD~\cite{cimpoi14describing}, and GTSRB~\cite{houben2013detection}, of which results are provided in Appendix \ref{appendix:more_datasets}. Our study employs two distinct model architectures: Preact-ResNet18~\cite{he2016identity} and VGG19-BN~\cite{simonyan2014very}. The results of VGG19-BN are provided in Appendix \ref{appendix:vgg19_bn}.

\paragraph{Evaluation metrics.} In this work, the metrics evaluating the performance of backdoor attacks are Accuracy (ACC) and Attack Success Rate (ASR). The metrics used by the defender are True Positive Rate (TPR), False Positive Rate (FPR), and F1 score. In the tables presenting our results, the top performer is highlighted in \textbf{bold}, and the runner-up is marked with an \underline{underline}.

\input{table/agpd_preact_fscore}

\subsection{Detection effectiveness evaluation}

\paragraph{All-to-one \& all-to-all attacks.}

Tab. \ref{tab:compared_results} showcases the detection performance of AGPD with eight compared methods against 12 backdoor attacks on Preact-ResNet18. For all-to-one attacks and all-to-all attacks, AGPD can achieve averaged TPR of 96.66\% on the CIFAR-10 and 99.92\% on the Tiny ImageNet, exceeding the runner-up by 18.23\% and 11.8\% respectively. 

The averaged FPR of AGPD not only ranks within the top-2 lowest among all detection methods but also approaches a near $0\%$ level.
Meanwhile, its average F1 score is 12.71\% higher than that of the second-best method.

For the \textbf{activation-based} methods, like Beatrix, SCAn and Spectral, we find that they effectively identifies the majority of poisoned samples in attacks where poisoned and clean samples are separated in the activation space. However, its performance deteriorates when this separation is not present, such as CTRL (see t-SNE results in Appendix \ref{appendix:tsne}). The failure of AC could be caused by the high poisoning ratio. For \textbf{input-based} method like STRIP, they exhibit low TPRs in attacks such as WaNet and Input-Aware. This underperformance is likely because the perturbed inputs generated by a poisoned sample also display high entropy in their predictions, similar to those of a clean sample, thereby complicating the distinction between them. CD shows relatively good detection effectiveness across most attacks with an average TPR of 76.76\%, although it also suffers from higher FPRs. The reason could be that the masks derived from cognitive distillation for poisoned and clean samples are too similar under $L_1$ norm, leading to misclassification of some clean samples as poisoned. For \textbf{loss-based} method like ABL, they perform well in attacks with attacks such as BadNets, Blended, SIG, and CTRL. However, their effectiveness decreases when facing attacks with dynamic triggers, such as WaNet, Input-Aware, and Adap-Blend. These attacks require more training epochs for models to learn the connection between trigger and target label, which means that the loss of poisoned samples does not significantly decrease in the early epochs \cite{wu2022backdoorbench}. Regarding the ASSET method, we observed potential impacts on detection performance due to differences in the model used compared to the original work. Thus, we provide the results of ASSET on ResNet18 in Appendix \ref{appendix:asset}.

\input{table/multi-target-preact-fscore}

\paragraph{Multi-target attacks.}

Tab. \ref{tab:multi_preact} summarizes the performance of AGPD and the compared methods in a multi-target attack scenario. In our experiment setting, $\{5,6,7,8,9\}$ are chosen as the source class and the target labels are set to $t = (y+C) \mod K$, where $C$ equals 5.

Single-trigger attack and multi-trigger attack are two categories of multi-target attack. In single-trigger attacks, the same trigger injected into samples from different source classes is classified into their corresponding target classes. In the multi-trigger attack, we use five triggers from different backdoor attacks (\ie BadNets, Blended, LF, SSBA, and SIG), and each trigger added to the samples in the corresponding class will be classified into its designated target class. We observed that for activation-based methods, the multi-trigger attack poses a greater challenge than single-trigger attacks, whereas loss-based methods seem more robust against multi-trigger attacks. Additionally, the failure of SCAn might be caused by their anomaly detection for target class(es) is not effective in the multi-target attacks. However, compared with baseline methods, AGPD achieves good performance in both single-trigger and multi-trigger attacks, with averages of TPR, FPR, and F1 score at 97.42\%, 0.06\%, and 98.41\%, respectively.

\begin{figure}[htbp]
    \centering
    \includegraphics[width=0.8\linewidth]{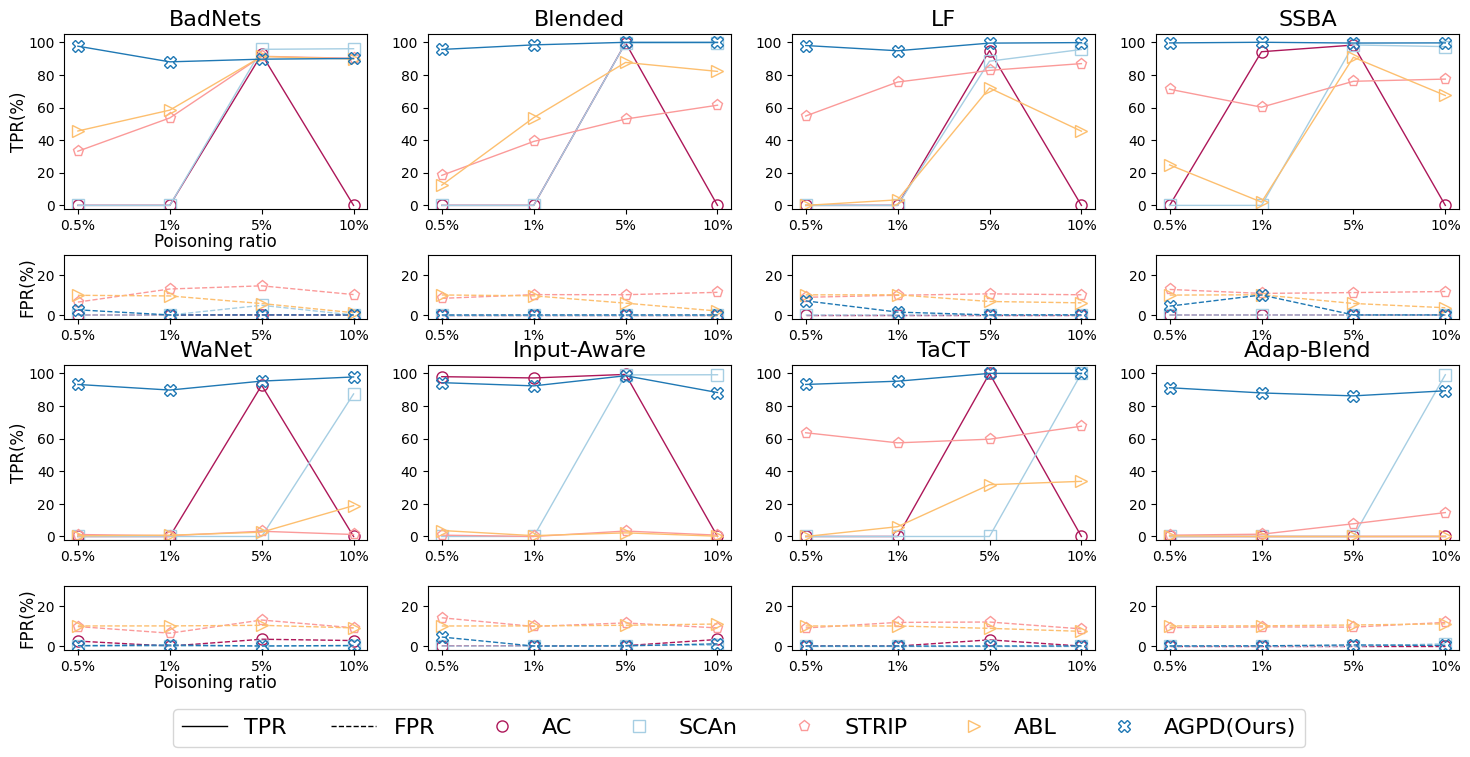}
    \caption{Detection performance of AGPD and the compared detectors with poisoing ratios ranging from $0.5\%$ to $10\%$.}
    \label{fig:ratio}
\vspace{-1.7em}
\end{figure}

\paragraph{Different poisoning ratios.}

We evaluated the detection effectiveness of AGPD and four other methods, chosen from activation-based, input-based, and loss-based detection methods, under varying poisoning ratios. There are four poisoning ratios used: $\{0.5\%, 1\%, 5\%, 10\%\}$, covering a range from low to high poisoning ratios. As illustrated in Fig.~\ref{fig:ratio}, the performance of most detectors is notably influenced by the poisoning ratio. Particularly, a low poisoning ratio (\eg 0.5\%) presents a substantial challenge for most detectors, with the TPRs of AC and SCAn almost nearing zero. However, our method can achieve good performance in this situation, with TPR around 90\% and FPR lower than that of other detectors under most attacks. And with the poisoning ratio increased, the performance of our method is still stable.

\subsection{Analysis}

\paragraph{Activation vs. activation gradient.}

\begin{wraptable}{r}{0.42\textwidth}
\centering
\caption{Silhouette scores of the target class under eight attacks, measured in activation and activation gradient spaces, using Preact-ResNet18.}
\scalebox{0.52}{
\begin{tabular}{ccccc}
\toprule
                    & BadNets & Blended & LF    & SSBA   \\
                    \cmidrule{2-5}
Activation          & 0.529   & 0.485   & 0.472 & 0.497 \\
Activation Gradient & 0.664   & 0.696   & 0.610 & 0.623 \\
\midrule
\midrule
& WaNet & Input-Aware & TaCT  & Adap-Blend \\
\cmidrule{2-5}
Activation& 0.544 & 0.457       & 0.379 & 0.403      \\
Activation Gradient& 0.605 & 0.466       & 0.515 & 0.491     \\

\bottomrule
\end{tabular}
}
\label{tab:s_score}
\end{wraptable}

To demonstrate that the separations of clean and poisoned samples differ in activation gradient and activation spaces, we show the distribution of cosine similarities between samples and the clean basis from both spaces. Due to the space limitation, we provide the results in Appendix \ref{appendix:separation}. To quantify the separation, we utilize Silhouette Score~\cite{rousseeuw1987silhouettes} to measure the distance between two clusters in both spaces across all convolutional layers. The range of the Silhouette Score is between -1 and 1, with higher values indicating better separability between the two clusters.  

Considering that different depths of convolutional layers correspond to different separations, Tab. \ref{tab:s_score} presents the maximum Silhouette Scores among all convolutional layers for eight backdoor attacks, from which we find that Silhouette Scores are larger in activation gradient space.

\vspace{-0.8em}

\begin{figure}[h]
\vspace{-0.5em}
    \centering
    \begin{subfigure}[b]{0.48\textwidth}
        \centering
        \includegraphics[width=\textwidth]{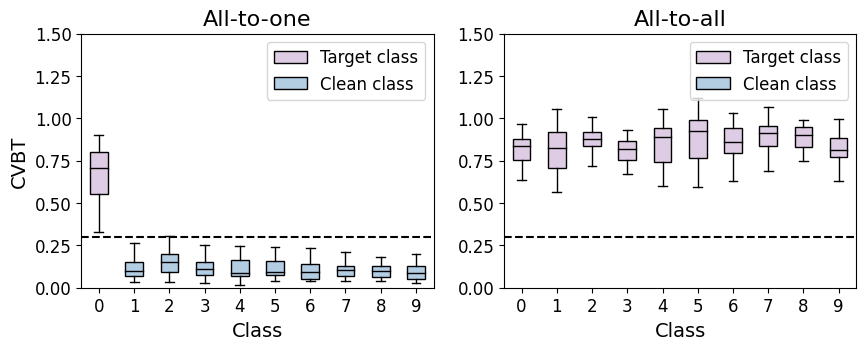}
        \vspace{-1.4em}
        \caption{}
        \label{fig:cvbt}
    \end{subfigure}
    \begin{subfigure}[b]{0.25\textwidth}
        \centering
        \includegraphics[width=\textwidth]{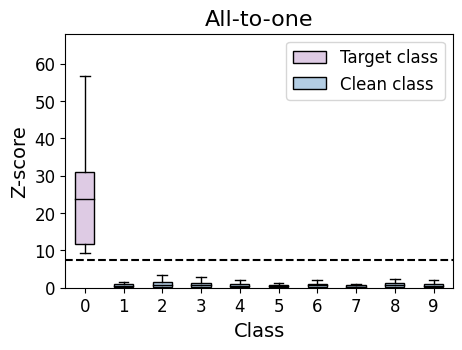}
        \vspace{-1.4em}
        \caption{}
        \label{fig:z_score}
    \end{subfigure}
    \begin{subfigure}[b]{0.25\textwidth}
        \centering
        \includegraphics[width=\textwidth]{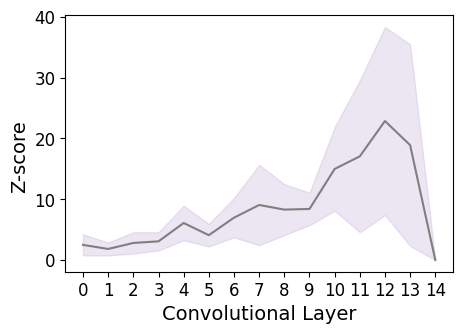}
        \vspace{-1.4em}
        \caption{}
        \label{fig:z_score_layer}
    \end{subfigure}
    \vspace{-.4em}
    \caption{Statistical analysis of $\rho$ and $z$ across classes and convolutional layers using the CIFAR-10 and Preact-ResNet18. \textbf{(a)} $\rho$ values for all classes in both all-to-one and all-to-all attacks. \textbf{(b)} $z$ for all-to-one attacks. \textbf{(c)} Mean and standard deviation of the maximum $z$ across all layers in multiple backdoored models.}
    \label{fig:statistic}
\vspace{-1.0em}
\end{figure}

\paragraph{Statistic of metrics.} In Fig. \ref{fig:statistic}, we present the statistical results of CVBT metric ($\rho$) and its Z-score ($z$) for both all-to-one and all-to-all attacks. In the left of Fig. \ref{fig:cvbt}, $\rho$ of the target class are significantly higher than those for clean classes for both attacks. 
Since clean classes are absent in all-to-all attacks, using $z$ to detect the outliers for identifying the target class could be ineffective. Therefore, if the value of $z$ fails to identify the target class, we utilize the threshold defined by the boundary, which serves as a boundary between the target and clean classes, to determine if a class is the target class. It is set to 0.3 and depicted as a dashed line in Fig. \ref{fig:statistic}
This is validated in Fig. \ref{fig:cvbt} right image, where $\rho$ values for the target classes in all-to-all attacks exceed our boundary. Fig. \ref{fig:z_score} displays the distribution of $z$ for the target and clean classes in the all-to-one attacks, clearly separated by the dashed line, which represents the threshold when identifying the target class. Moreover, Fig. \ref{fig:z_score_layer} illustrates the mean and standard deviation curves of $z$ of target classes across different convolutional layers of the model. We observe that $z$ tend to be higher in the later intermediate convolutional layers, indicating a stronger separation activation gradients between poisoned and clean samples at these layers.

\paragraph{Accuracy of target label identification.}

We compare the accuracy of target class identification of AGPD with other three detection methods which are Beatrix~\cite{mabeatrix}, SCAn~\cite{tang2021demon}, and NC~\cite{wang2019neural}. To evaluate their performance, we trained 120 backdoor models on CIFAR-10. The attack methods contain 8 non-clean label backdoor attacks, where the poisoning ratio ranges from $1\%$ to $10\%$, and the target label is from 0 to 4. The results of detection accuracy are shown in ~\cref{fig:target_acc}. Note that the accuracy of target class identification of AGPD is higher than the compared method under different poisoning ratios. Even though SCAn has a competitive performance with AGPD, the accuracy of SCAn is lower than 20\% with 1\% poisoning ratio. 

\subsection{Sensitivity test}
\paragraph{Influence of number of clean samples.}
In this part, we explore the influence of the size of the additional clean dataset on the detection performance of AGPD. We also consider the scenario that the additional dataset collected by the defender is out of distribution (OOD). 

\begin{figure}[h]
    \centering
    \begin{subfigure}[b]{0.3\textwidth}
        \centering
        \includegraphics[width=\textwidth]{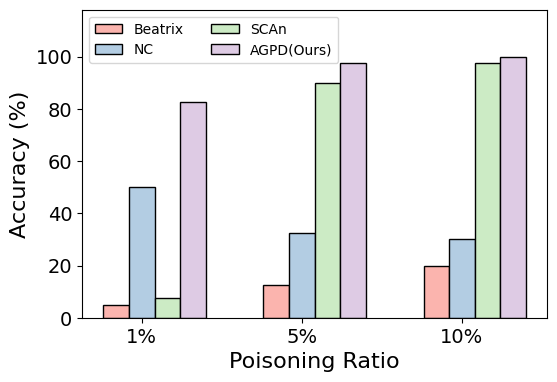}
        \captionsetup{labelformat=empty}
        \caption{}
        \label{fig:target_acc}
    \end{subfigure}
    \begin{subfigure}[b]{0.34\textwidth}
        \centering
        \includegraphics[width=\textwidth]{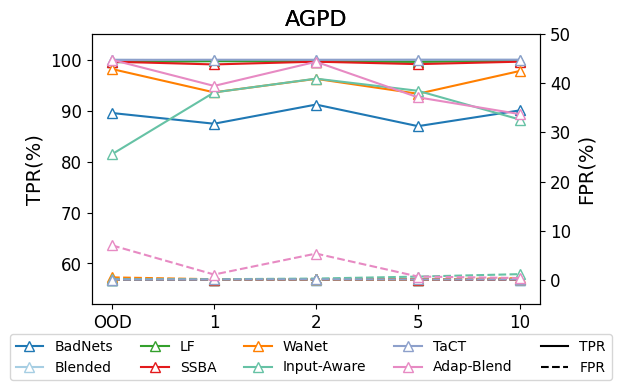}
        \captionsetup{labelformat=empty}
        \caption{}
        \label{fig:number}
    \end{subfigure}
    \begin{subfigure}[b]{0.31\textwidth}
        \centering
        \includegraphics[width=\textwidth]{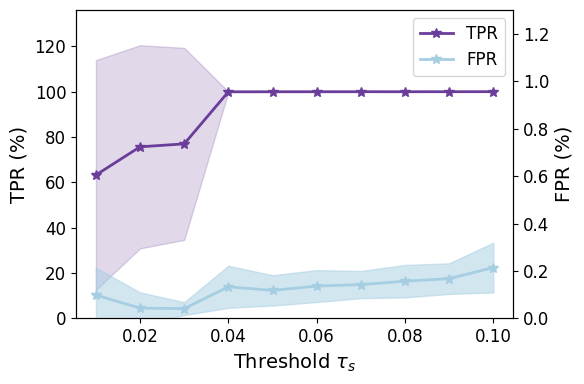}
        \captionsetup{labelformat=empty}
        \caption{}
        \label{fig:xi}
    \end{subfigure}
    \caption{\textbf{left:} Accuracy of AGPD on identifying target class(es) compared with three compared methods. \textbf{middle:} Detection performance of AGPD with varying numbers of clean samples. \textbf{right:} Means and standard deviations of TPR and FPR at different threshold $\tau_s$.}
    \label{fig:combine}
\end{figure}

We collect the OOD dataset of CIFAR-10 from the same 10 classes of CIFAR-5m~\cite{cifar5m}, and we extract 10 samples from each class. 
The additional clean dataset which is in distribution (ID) is collected from the test dataset. ~\cref{fig:number} shows the results of our method with different sizes of the additional clean dataset.
We found that a large number of clean samples can help AGPD decrease FPR close to zero. However, AGPD can still achieve high TPR even in extreme cases, such as one sample per class or even OOD samples.

When the number of clean samples in each class is one, the TPR values of AGPD on many attacks are above 90\%. In summary, our method necessitates a smaller additional clean dataset.

\paragraph{Influence of threshold $\tau_s$.}

In the poisoned sample filtering stage, we aim to eliminate samples scoring below $\tau_s$ at each iteration until no samples remain in the target class. To better understand the impact of the threshold value $\tau_s$, we designed an experiment where we varied $\tau_s$ from 0.01 to 0.1 and recorded their TPRs and FPRs. According to Fig. \ref{fig:xi}, the TPR of AGPD is relatively low with significant variability at smaller $\tau_s$ values, yet it stabilizes at $100\%$ with increasing tau, while the FPR remains consistently low throughout the variation of tau.
Moreover, it can ensure stable detection performance of AGPD across a broad range of values. 

%% file: table/agpd_preact_fscore.tex
\begin{table}[h]
\vspace{-.8em}
\centering
\caption{The detection performance of AGPD and compared detectors on CIFAR-10 and Tiny ImageNet. The results are evaluated on Preact-ResNet18.}
\renewcommand\arraystretch{1.3}
\setlength{\tabcolsep}{2pt} 
\scalebox{0.45}{
\begin{tabular}{c|c|c|ccc|ccc|ccc|ccc|ccc|ccc|ccc|ccc|ccc}

\toprule
\multirow{2}{*}{Dataset}       & \multirow{2}{*}{Attack} & Backdoored  & \multicolumn{3}{c|}{AC~\cite{ac}} & \multicolumn{3}{c|}{Beatrix~\cite{mabeatrix}} & \multicolumn{3}{c|}{SCAn~\cite{tang2021demon}} & \multicolumn{3}{c|}{Spectral~\cite{tran2018spectral}} & \multicolumn{3}{c|}{STRIP~\cite{gao2019strip}} & \multicolumn{3}{c|}{ABL~\cite{li2021anti}} & \multicolumn{3}{c|}{CD~\cite{huang2023distilling}} & \multicolumn{3}{c|}{ASSET~\cite{pan2023asset}} & \multicolumn{2}{c}{AGPD} \\

&                         & ACC/ASR     & TPR$\uparrow$     & FPR$\downarrow$    & F1$\uparrow$ & TPR$\uparrow$     & FPR$\downarrow$      & F1$\uparrow$ & TPR$\uparrow$      & FPR$\downarrow$    & F1$\uparrow$ & TPR$\uparrow$     & FPR$\downarrow$     & F1$\uparrow$ & TPR$\uparrow$      & FPR$\downarrow$     & F1$\uparrow$ & TPR$\uparrow$     & FPR$\downarrow$     & F1$\uparrow$ & TPR$\uparrow$     & FPR$\downarrow$     & F1$\uparrow$ & TPR$\uparrow$     & FPR$\downarrow$     & F1$\uparrow$ & TPR$\uparrow$      & FPR$\downarrow$    & F1$\uparrow$ \\
\midrule
\multirow{13}{*}{\rotatebox{90}{CIFAR-10}}&BadNets \cite{gu2019badnets}&91.82/93.79&0.00&\textbf{0.00}&0.00&87.24&8.95&65.15&\textbf{96.04}&\textbf{0.00}&\textbf{97.98}&16.76&1.30&26.09&\underline{90.16}&10.19&63.97&89.74&1.14&89.74&78.02&43.87&27.24&3.16&47.66&1.19&90.06&\underline{0.03}&\underline{94.65}\\
&Blended \cite{chen2017targeted}&93.69/99.75&0.00&\textbf{0.00}&0.00&47.60&5.07&49.28&\underline{99.62}&\textbf{0.00}&\underline{99.81}&28.04&0.05&43.64&61.42&11.31&46.67&82.14&1.98&82.14&85.06&49.62&26.93&3.70&12.66&3.40&\textbf{99.98}&\underline{0.02}&\textbf{99.88}\\
&LF \cite{zeng2021rethinking}&93.01/99.05&0.00&\textbf{0.00}&0.00&0.00&10.72&0.00&\underline{95.58}&\underline{0.01}&\underline{97.71}&0.04&3.16&0.06&86.92&10.09&62.59&45.48&6.06&45.48&88.44&24.33&43.42&3.80&10.28&3.87&\textbf{99.80}&0.07&\textbf{99.60}\\
&SSBA \cite{li2021invisible}&92.88/97.06&0.00&\textbf{0.00}&0.00&10.26&8.54&10.97&\underline{97.34}&\underline{0.01}&\underline{98.60}&27.14&0.15&42.24&77.42&11.71&54.75&67.38&3.62&67.38&91.30&3.76&81.12&3.56&46.36&1.37&\textbf{99.62}&0.04&\textbf{99.63}\\
&SIG \cite{SIG}&93.40/95.43&0.00&\textbf{0.00}&0.00&0.00&\textbf{0.00}&0.00&\underline{99.52}&\textbf{0.00}&\textbf{99.74}&0.00&1.58&0.00&99.44&9.68&51.88&90.80&5.75&60.53&85.88&21.28&45.51&0.48&32.73&0.13&\textbf{100.0}&\underline{0.04}&\underline{99.66}\\
&CTRL&95.52/98.8&0.00&9.92&0.00&0.00&6.26&0.00&0.00&5.00&0.00&0.40&15.77&0.20&\textbf{99.80}&9.47&52.57&90.32&5.77&60.21&99.44&\underline{0.55}&\underline{97.31}&67.88&51.60&11.82&\underline{99.76}&\textbf{0.01}&\textbf{99.78}\\
&WaNet \cite{nguyen2021wanet}&89.68/96.94&0.00&2.77&0.00&0.92&9.74&0.94&\underline{87.39}&\textbf{0.07}&\underline{92.96}&0.90&2.97&1.38&1.22&9.13&1.28&18.92&9.08&18.31&86.88&79.77&19.20&0.55&1.22&1.01&\textbf{97.80}&\underline{0.31}&\textbf{97.40}\\
&\small{Input-Aware~\cite{nguyen2020input}}&90.82/98.17&0.00&3.31&0.00&0.41&10.92&0.39&\textbf{99.15}&\textbf{0.47}&\textbf{97.36}&1.51&2.90&2.34&0.81&9.07&0.86&0.17&11.02&0.17&82.85&18.99&46.84&2.82&60.13&0.90&\underline{88.25}&\underline{1.13}&\underline{88.61}\\
&TaCT \cite{tang2021demon}&93.21/95.95&0.00&\textbf{0.00}&0.00&75.94&19.98&42.69&\textbf{100.0}&\textbf{0.00}&\textbf{99.99}&29.76&\underline{0.03}&45.78&67.60&8.59&55.20&33.80&7.36&33.80&\underline{80.52}&53.90&24.19&7.74&59.90&2.64&\textbf{100.0}&0.07&\underline{99.68}\\
&\small{Adap-Blend \cite{adapt}}&92.87/66.17&0.00&\textbf{0.00}&0.00&4.62&8.33&5.14&\textbf{99.16}&1.15&\textbf{94.66}&24.34&0.47&37.85&14.66&11.82&13.27&0.08&11.10&0.00&0.00&\textbf{0.00}&0.00&\underline{97.22}&39.52&37.88&89.32&\underline{0.33}&\underline{92.89}\\
&BadNets-A2A&91.93/74.40&28.76&4.51&33.96&40.28&9.25&\underline{36.04}&0.00&\textbf{0.00}&0.00&0.00&1.67&0.00&1.80&17.08&1.41&2.48&10.84&2.48&\underline{67.20}&45.72&23.23&3.32&3.31&4.99&\textbf{97.32}&\underline{0.02}&\textbf{98.57}\\
&SSBA-A2A&93.46/87.84&50.02&2.66&\underline{57.51}&19.04&5.26&22.88&0.00&\textbf{0.00}&0.00&0.00&1.67&0.00&12.74&9.87&12.64&0.08&11.00&0.96&\underline{75.54}&44.43&26.26&48.74&49.56&16.39&\textbf{98.06}&\underline{0.02}&\textbf{98.95}\\
\rowcolor{gray!20}
&Avg.&&6.57&1.93&7.62&23.86&8.58&19.46&72.82&\underline{0.56}&\underline{73.23}&10.74&2.64&16.63&51.17&10.67&34.76&43.45&7.06&38.43&\underline{76.76}&32.18&38.44&20.25&34.58&7.13&\textbf{96.66}&\textbf{0.17}&\textbf{97.44}\\
\midrule
\midrule
\multirow{9}{*}{\rotatebox{90}{Tiny ImageNet}}&BadNets \cite{gu2019badnets}&56.12/99.90&0.00&0.40&0.00&1.11&9.64&1.18&\textbf{100.0}&\textbf{0.00}&\textbf{100.0}&14.04&0.18&24.28&\textbf{100.0}&11.51&65.89&95.49&0.50&95.49&66.91&51.31&21.28&95.11&38.62&35.05&\underline{99.90}&\underline{0.16}&\underline{99.24}\\
&Blended \cite{chen2017targeted}&55.53/97.57&0.00&1.26&0.00&0.53&10.06&0.55&\underline{99.85}&\textbf{0.00}&\textbf{99.92}&11.45&0.47&19.80&96.51&11.89&63.59&90.18&1.09&90.18&93.29&9.94&66.00&78.53&60.74&21.66&\textbf{100.0}&\underline{0.05}&\underline{99.78}\\
&LF \cite{zeng2021rethinking}&55.21/98.51&15.05&1.26&23.82&19.36&9.15&19.20&63.86&\textbf{0.00}&77.94&11.35&0.48&19.62&85.97&9.72&62.88&87.44&1.40&\underline{87.44}&\underline{95.22}&6.65&74.65&54.28&50.69&17.78&\textbf{100.0}&\underline{0.10}&\textbf{99.56}\\
&SSBA \cite{li2021invisible}&55.97/97.69&0.00&1.05&0.00&0.45&8.70&0.50&59.11&\textbf{0.00}&74.30&13.97&0.19&24.15&\textbf{99.96}&11.20&66.46&95.24&0.53&\underline{95.24}&88.07&20.94&46.78&26.90&57.36&8.37&\underline{99.89}&\underline{0.04}&\textbf{99.75}\\
&WaNet \cite{nguyen2021wanet}&58.33/90.35&13.73&0.80&22.60&75.94&11.88&52.23&62.72&\textbf{0.00}&77.09&11.37&0.45&19.65&6.38&11.11&5.97&\underline{94.40}&1.27&\underline{91.35}&76.73&20.18&42.83&0.00&15.12&0.00&\textbf{99.77}&\underline{0.12}&\textbf{99.30}\\
&\small{Input-Aware~\cite{nguyen2020input}}&57.5/99.75&0.00&0.68&0.00&64.97&10.30&49.13&\underline{99.65}&\textbf{0.00}&\textbf{99.82}&12.92&0.29&22.32&12.02&11.08&10.97&72.52&3.53&70.18&80.22&28.93&36.41&3.07&5.97&3.97&\textbf{99.89}&\underline{0.04}&\underline{99.73}\\
&TaCT \cite{tang2021demon}&54.93/91.25&0.00&1.02&0.00&45.51&10.13&38.45&\textbf{100.0}&\textbf{0.00}&\textbf{100.0}&15.48&\underline{0.03}&26.75&80.03&16.37&48.90&32.25&7.53&32.25&99.58&99.45&18.19&35.77&56.31&12.20&\underline{99.99}&0.43&\underline{98.11}\\
&\small{Adap-Blend \cite{adapt}}&54.55/96.35&0.00&0.82&0.00&9.56&9.39&9.85&47.24&\textbf{0.05}&\underline{63.98}&15.14&\underline{0.06}&26.18&\underline{77.73}&15.85&48.52&8.97&10.11&8.97&73.69&39.65&27.77&71.95&49.92&25.19&\textbf{99.96}&0.17&\textbf{99.24}\\
\rowcolor{gray!20}
&Avg.&&3.60&0.91&5.80&27.18&9.91&21.39&79.05&\textbf{0.01}&\underline{86.63}&13.21&0.27&22.84&69.83&12.34&46.65&72.06&3.24&71.39&\underline{84.21}&34.63&41.74&45.70&41.84&15.53&\textbf{99.92}&\underline{0.14}&\textbf{99.34}\\
\bottomrule
\end{tabular}
}
\vspace{-.8em}
\label{tab:compared_results}
\end{table}

%% file: table/multi-target-preact-fscore.tex
\begin{table}[h]
\vspace{-0.6em}
\centering
\caption{The detection performance of AGPD and the compared methods against multi-target attacks on CIFAR-10. The model structure is Preact-ResNet18.}
\renewcommand\arraystretch{1.3}
\setlength{\tabcolsep}{2pt} 
\scalebox{0.42}{
\begin{tabular}
{c|c|ccc|ccc|ccc|ccc|ccc|ccc|ccc|ccc|ccc}
\toprule
\multirow{2}{*}{Attack}  & Backdoored  & \multicolumn{3}{c|}{AC~\cite{ac}} & \multicolumn{3}{c|}{Beatrix~\cite{mabeatrix}} & \multicolumn{3}{c|}{SCAn~\cite{tang2021demon}} & \multicolumn{3}{c|}{Spectral~\cite{tran2018spectral}} & \multicolumn{3}{c|}{STRIP~\cite{gao2019strip}} & \multicolumn{3}{c|}{ABL~\cite{li2021anti}} & \multicolumn{3}{c|}{CD~\cite{huang2023distilling}} & \multicolumn{3}{c|}{ASSET~\cite{pan2023asset}} & \multicolumn{2}{c}{AGPD} \\

& ACC/ASR     & TPR$\uparrow$     & FPR$\downarrow$    & F1$\uparrow$ & TPR$\uparrow$     & FPR$\downarrow$      & F1$\uparrow$ & TPR$\uparrow$      & FPR$\downarrow$    & F1$\uparrow$ & TPR$\uparrow$     & FPR$\downarrow$     & F1$\uparrow$ & TPR$\uparrow$      & FPR$\downarrow$     & F1$\uparrow$ & TPR$\uparrow$     & FPR$\downarrow$     & F1$\uparrow$ & TPR$\uparrow$     & FPR$\downarrow$     & F1$\uparrow$ & TPR$\uparrow$     & FPR$\downarrow$     & F1$\uparrow$ & TPR$\uparrow$      & FPR$\downarrow$    & F1$\uparrow$ \\
 \hline
\textbf{Single-trigger} & &  &  &  &  &  & & & &  &  &  & &  &  & &  & &  &&&&&&&&& \\
BadNets&91.38/80.34&\underline{97.46}&0.70&\underline{95.64}&19.38&0.96&30.28&0.00&\textbf{0.00}&0.00&5.42&16.06&4.34&2.78&13.01&2.53&1.22&10.98&1.22&47.42&31.62&21.95&22.84&20.79&14.74&\textbf{98.50}&\underline{0.08}&\textbf{98.89}\\
Blended&93.57/91.60&\underline{79.72}&0.25&\underline{87.61}&11.76&4.34&15.59&0.00&\textbf{0.00}&0.00&14.90&15.01&11.92&0.18&6.49&0.23&2.08&10.88&2.08&74.66&55.91&22.03&19.64&8.70&19.85&\textbf{98.48}&\underline{0.01}&\textbf{99.20}\\
LF&93.54/93.82&\textbf{99.72}&0.03&\textbf{99.71}&20.40&2.33&28.86&0.00&\textbf{0.00}&0.00&33.78&12.91&27.02&5.32&7.88&6.04&1.82&10.91&1.82&62.02&41.63&23.11&0.24&0.66&0.45&\underline{99.20}&\underline{0.01}&\underline{99.55}\\
SSBA&93.28/92.38&\textbf{99.44}&0.04&\textbf{99.52}&4.50&0.79&8.07&0.00&\textbf{0.00}&0.00&32.36&13.07&25.89&48.26&15.65&33.39&11.48&9.84&11.48&72.56&25.16&36.37&13.52&8.02&14.56&\underline{98.92}&\underline{0.02}&\underline{99.36}\\

\midrule
\textbf{Multi-triggers} & & & &  &  & & &  &  &  &  &  &  & & & & & & &&&&&&&&&\\
\scalebox{0.8}{BadNets+Blended+LF+SSBA+SIG}&91.62/92.10&58.06&\underline{0.08}&\underline{73.15}&2.92&2.60&4.62&0.00&\textbf{0.00}&0.00&14.64&15.05&11.71&22.72&13.27&18.77&50.10&5.54&50.10&\underline{63.20}&18.60&38.23&56.86&54.68&17.52&\textbf{92.02}&0.18&\textbf{95.06}\\

\midrule
\rowcolor{gray!20}
Avg.&&\underline{86.88}&0.22&\underline{91.13}&11.79&2.20&17.48&0.00&\textbf{0.00}&0.00&20.22&14.42&16.18&15.85&11.26&12.19&13.34&9.63&13.34&63.97&34.58&28.34&22.62&18.57&13.42&\textbf{97.42}&\underline{0.06}&\textbf{98.41}\\

\bottomrule
\end{tabular}
}
\vspace{-1.2em}
\label{tab:multi_preact}
\end{table}

%% file: section/conclusion.tex
\section{Conclusion}
In this paper, we introduce a novel perspective of gradient circular distribution (GCD). Based on GCD, we observe that the dispersion of GCD of target class is larger, and poisoned samples are separated from clean ones. Inspired by the observation, we propose two practical metrics and design a novel detection method, AGPD. Our experiments demonstrate that this method successfully identifies target class(es) under various backdoor attack scenarios, including all-to-one, all-to-all, and multi-target attacks. Extensive experimental results show that our method achieves good performance on the task of poisoned sample detection.

\paragraph{Limitation and future work.} 
The proposed method involves scanning through all layers of the model to identify the one where poisoned and clean samples are most distinctly separated. The computational cost may be high if applying it on large scale datasets, which may require training large models with multiple layers to extract activation gradients. 
However, as analyzed in Fig. \ref{fig:statistic}-right, the activation gradient directions of poisoned and clean samples typically exhibit greater separation in higher layers of the model. 
Inspired, in future we plan to narrow the range of selected layers to further enhance the efficiency of the proposed method. 

%% file: section/appendix.tex
\appendix
\section{In summary}
There are additional materials presented in the Appendix, including: \textbf{(1)} details of datasets and models we used; \textbf{(2)} hyperparameter settings in model training; \textbf{(3)} computational overhead of the proposed method and the compared methods; \textbf{(4)} Details of algorithm and stopping criteria; \textbf{(5)} performance of AGPD on more datasets; \textbf{(6)} separation of poisoned and clean samples in activation and activation gradient space; \textbf{(7)} more GCDs examples; \textbf{(8)} The results of the compared ASSET on ResNet18. \textbf{(9)} t-SNE results of ten attacks on Preact-ResNet18.

\section{Datasets and models}
We evaluate the performance of AGPD on five popular datasets and two model structures. In main paper, we have provide the results of two datasets, including CIFAR-10 \cite{krizhevsky2009learning} and Tiny ImageNet \cite{le2015tiny}. Besides, we extend our evaluations to the datasets which are closer to real-world scenarios, such as ImageNet(subset)-200 \cite{deng2009imagenet}, the Textures dataset DTD \cite{cimpoi14describing}, and the traffic signs dataset GTSRB \cite{houben2013detection}. The results of the these datasets are provided in Appendix \ref{appendix:more_datasets}. The details of all datasets are illustrated in Tab. \ref{tab:data_info} and Tab. \ref{tab:model_structure}. 

\begin{table}[h]
    \centering
    \caption{The information about five datasets.}
    \renewcommand\arraystretch{1.3}
    \setlength{\tabcolsep}{3pt}
    \scalebox{0.8}{
    \begin{tabular}{c|c|c|c|c}
    \toprule
        Dataset &  Categories & Image size & Training samples & Testing samples\\
        \midrule
        CIFAR-10 & 10 & $32\times 32$ & 50,000 & 10,000 \\
        Tiny ImageNet & 200 & $64 \times 64$ & 90,000 & 10,000 \\
        ImageNet(subset200) & 200 & $224\times 224$ & 90,000 & 10,000\\
        GTSRB  & 43 & $32\times 32$ & 39,209 & 12,630 \\
        DTD & 47 & $224\times 224$ & 3,760 & 1,880\\
    \bottomrule

    \end{tabular}
}
    \label{tab:data_info}
\end{table}

\input{table/appendix/model_structure}

\section{Hyperparameter settings}
\label{appendix:hyper}
In this part, we will introduce the hyperparameter settings of model training and backdoor attacks. 
\paragraph{Model training.} There are some common training hyperparameters across these attack methods, such as training epoch, learning rate, and optimizer. We display the setting of these common hyperparameters for each datasets in Tab. \ref{tab:hyper_training}.
\begin{table}[h]
    \caption{The common hyperparameters for training across five datasets.}
    \centering
    \renewcommand\arraystretch{1.3}
    \setlength{\tabcolsep}{3pt}
    \begin{tabular}{c|c|c|c|c}
    \toprule
        Dataset & Epoch & Learning rate & Batch size &Optimizer \\
    \midrule
        CIFAR-10 & 100 & 0.01 & 128 & SGD\\
        Tiny ImageNet & 200 & 0.01 & 128 & SGD \\
        ImageNet(subset200) & 200 & 0.1 &64& Adam \\
        GTSRB  & 50 & 0.01 &128& SGD \\
        DTD & 100 & 0.01 &64& SGD \\
    \bottomrule
    \end{tabular}
    \label{tab:hyper_training}
\end{table}

\begin{figure}[htbp]
  \centering
	\includegraphics[width=0.8\linewidth]{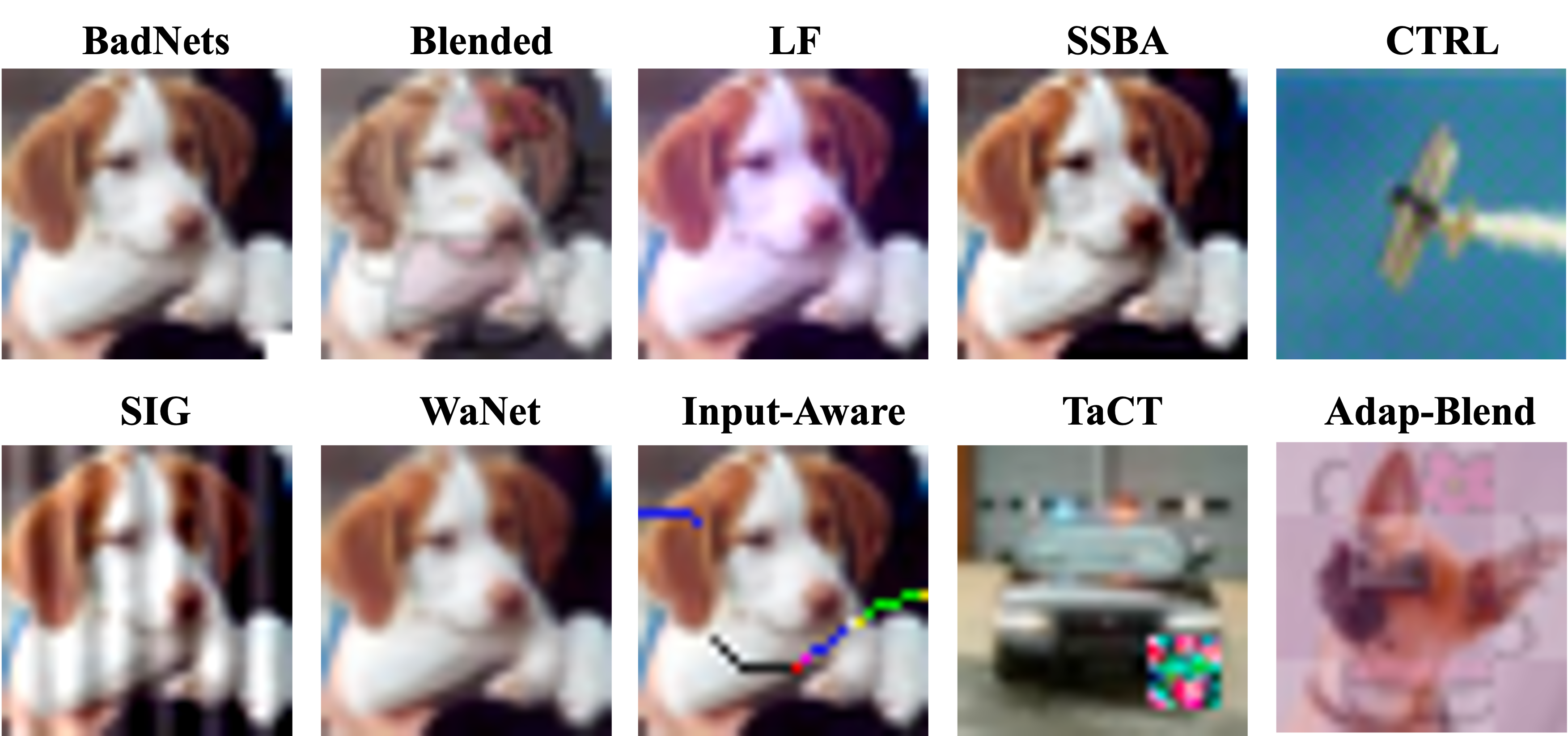}
   \caption{Examples of poisoned samples in various backdoor attacks.}
   \label{fig:poisoned samples}
\end{figure}

\paragraph{Backdoor attacks.}
The hyperparameters used in various backdoor attacks are listed in Tab. \ref{tab:hyper_attack}. For illustration purposes, we use CIFAR-10 as an example. If the attack does not have any specific hyper-parameters, we will denote this with `/'. We show the poisoned samples of various backdoor attacks in Fig.~\ref{fig:poisoned samples}. 

\input{table/appendix/hyper_attacks}

\section{Computation overhead}
Tab. \ref{tab:complexity} illustrates the computation overhead (RTX A5000 GPU) of AGPD and the compared detection method under eight backdoor attacks with 10\% poisoning ratio on CIFAR-10. We record the average of running time with standard deviation in bracket. 
\input{table/appendix/complexity_theo}

\newpage

\section{Details of algorithm and stopping criteria}
\label{appendix:sliding}

\subsection{The description of Algorithm}
\begin{figure}[h]
    \centering
    \includegraphics[width=0.5\linewidth]{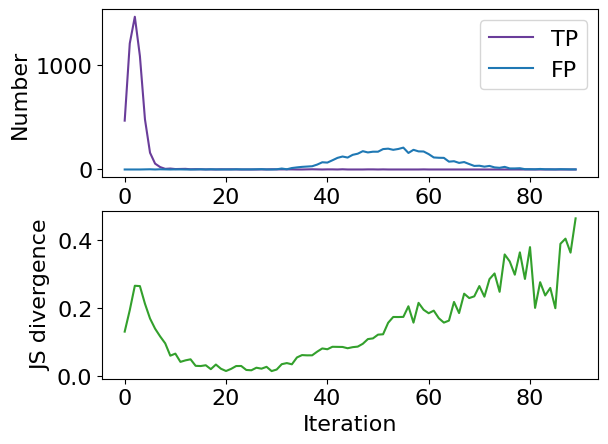}
    \caption{Trends of TP, FP, and JS divergence according to the iteration $t$.}
    \label{fig:js}
\end{figure}

The statement of the algorithm we used in Stage3 is described in Algorithm \ref{alg:psf}. An example of JS divergence across iterations is provided in Fig.~\ref{fig:js}. Assuming ground truth is known for samples in the target class, we can obtain True Positives (TP) and False Positives (FP) for each iteration. It can be observed that an optimal iteration exists where JS divergence is minimal and stabilizes. The rationale behind the trends in JS divergence is that in the early stages of filtering, the far-end basis is primarily updated by genuinely poisoned samples, effectively guiding the identification of such samples. As the process progresses into the middle stages, most poisoned samples have been filtered out, and the influence of the far-end basis on the remaining clean samples becomes minimal, resulting in little change in trust scores and small JS divergence. However, as the process extends into later stages, and more clean samples are inevitably filtered, the far-end basis is updated by these clean samples, leading to significant changes in the distribution of trustworthiness scores and an increase in JS divergence.

\input{alg/filtering}

\subsection{The stopping criteria}

To find an appropriate stopping iteration $ t^* $, we utilize the sliding window method to analyze the changes in JS divergence across all iterations. Our goal is to identify the iteration $ t $ where the JS divergence is minimal and stabilizes. Let the width of the window be $ w $, and let the JS divergence at each iteration $ t $ be $ JS(t) $, where $ t \in \{0, \dots, T-1\} $. The average $\mu_m$ and standard deviation $\sigma_m$ of each window starting at position $ m $ are defined as follows:

\begin{equation}
\left\{
\begin{aligned}
    \mu_m &= \frac{1}{w} \sum_{j=0}^{w-1} JS(m+j), \\
    \sigma_m &= \sqrt{\frac{1}{w} \sum_{j=0}^{w-1} \left( JS(m+j) - \mu_m \right)^2 }.
\end{aligned}
\right.
\end{equation}

Each window can be represented by a score $ S_m $ that combines the value of the average with the standard deviation. Since we are mainly focusing on stabilization, we design a metric as described in Eq. \ref{eq:window_score}, which amplifies the contribution of the standard deviation.

\begin{equation}
    S_m = \mu_m + \beta \sigma_m.
    \label{eq:window_score}
\end{equation}

After computing the score of all windows, we choose the window starting at $ m $ with the minimum score:
\begin{equation}
    m^* = \arg \min_m S_m.
    \label{eq:minimun_m}
\end{equation}
The stopping iteration is the middle of the optimal window, which is $ t^* = m^* + \frac{w}{2} $. In our method, we adopt $ w=10 $ and $ \beta=5 $.

\section{Evaluations on more datasets}
\label{appendix:more_datasets}
In this section, we present the results of AGPD and the compared methods on these datasets, including DTD \cite{cimpoi14describing}, GTSRB \cite{houben2013detection}, and ImageNet subset (200 classes) \cite{deng2009imagenet}. The datasets can be categorized into two types: balanced datasets (\eg DTD and ImageNet subset (200 classes)) and imbalanced datasets (\eg GTSRB). In balanced datasets, each category has the same number of samples. For example, DTD has 376 samples per class, and ImageNet-200 has 500 samples per class. In contrast, the number of samples per category in GTSRB varies from 210 to 2,250. Besides, We use Preact-ResNet18 \cite{he2016identity} as the model architecture when training on DTD and GTSRB, and adopt ResNet50 \cite{resnet50} for ImageNet subset (200 classes). Specifically, we compare AGPD with four detection methods: activation-based method (\eg SCAn \cite{tang2021demon}), input-based method (\eg STRIP \cite{gao2019strip}), and loss-based methods (\eg ABL \cite{li2021anti} and ASSET \cite{pan2023asset}) on DTD and GTSRB. The results are shown in Tab. \ref{tab:other_dataset}. Besides, the results of AGPD on ImageNetsubset (200 classes), are displayed in Tab. \ref{tab:imagenet}.

\input{table/appendix/dtd_gtsrb_result_fscore}

\input{table/appendix/imagenetsub}

As shown in Tables \ref{tab:other_dataset} and \ref{tab:imagenet}, AGPD achieves high performance on these datasets, with average TPRs of 99.61\%, 93.78\%, and 98.20\%, respectively. Meanwhile, the average FPRs are 1.43\%, 0.2\%, and 0.35\%. The experimental results demonstrate that our method is adaptable not only to balanced and imbalanced datasets but also to datasets with various image sizes.

Regarding the compared detection methods, we draw conclusions similar to those reported in our main paper. Specifically, SCAn fails when the distinction between poisoned and clean samples in the activation space is not obvious. STRIP struggles to effectively identify poisoned samples in the training dataset under attacks such as WaNet and Input-Aware. Similarly, ABL encounters challenges in achieving satisfactory detection performance against complex triggers, including those used in WaNet, Input-Aware, and Adap-Blend attacks.

\section{Evaluations on VGG19-BN}
\label{appendix:vgg19_bn}

\subsection{All-to-one \& all-to-all attacks.}
The results of VGG19-BN under all-to-one and all-to-all attacks are shown in Tab. \ref{tab:vgg_main}. It can be seen that even changing the model architecture, the detection performance of our method is still stable, achieving 96.46\% TPR on CIFAR-10 and 97.75\% on Tiny ImageNet, higher than that of the second-best 16.55\% and 7.89\%, respectively. Besides, the F1 score of AGPD are 90.90\% on CIFAR-10 and 98.36\% on Tiny ImageNet, exceeding the runner-up by 33.59\% and 16.27\%, respectively. The results indicate the dispersion of the activation gradients of poisoned samples and clean samples could exist across model structures, demonstrating the robust adapt ability of our method.

For the compared methods, we found that activation-based methods are significantly influenced by the model architecture. For instance, AC can identify a small portion of poisoned samples in all-to-one attacks with a 10\% poisoning ratio. Beatrix performs better on VGG19-BN, with the average TPR 20.78\% higher and the F1 score 17.39\% greater than on Preact-ResNet18. Conversely, SCAn struggles to detect a large number of poisoned samples in WaNet and Input-Aware attacks on the VGG19-BN.

\input{table/appendix/agpd_vgg19_fscore}

\newpage
\subsection{Performance of AGPD with VGG19-BN under various poisoning ratios.}
\begin{figure}[h]
    \centering
    \includegraphics[width=0.95\linewidth]{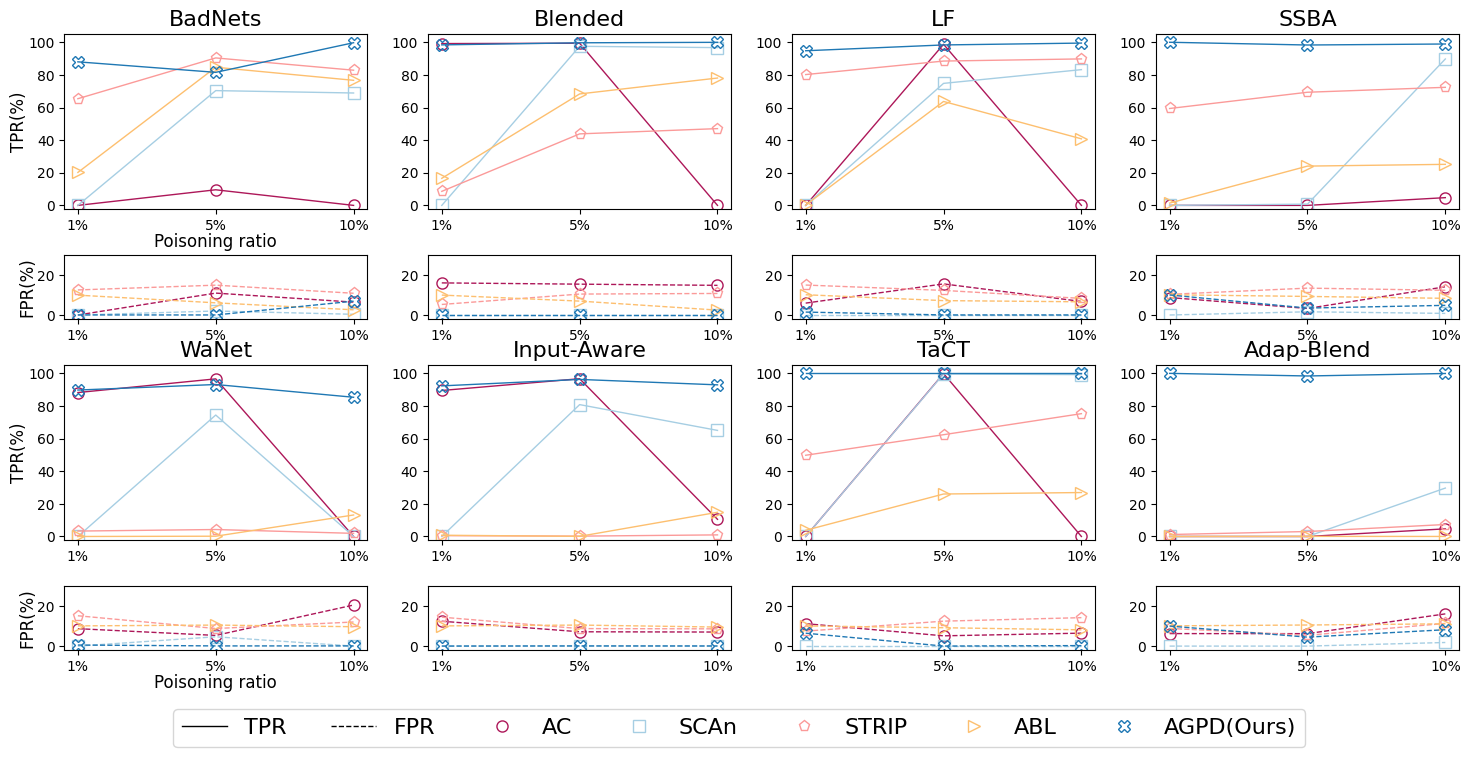}
    \caption{Detection performance of AGPD and the compared detectors with poisoning ratios ranging from $1\%$ to $10\%$.}
    \label{fig:ratio_vgg}
\end{figure}

We estimate the detection performance of AGPD against various backdoor attacks with different poisoning ratios and compare our method with four detectors. The results are displayed in Fig.~\ref{fig:ratio_vgg}. It can be seen that our method achieves a higher TPR under most attacks compared to other methods, which also maintain relatively low FPR.

\section{GCDs for various attacks on CIFAR-10}
\label{sec:gcds}
\begin{figure}[h]
  \centering
	\includegraphics[width=0.8\linewidth]{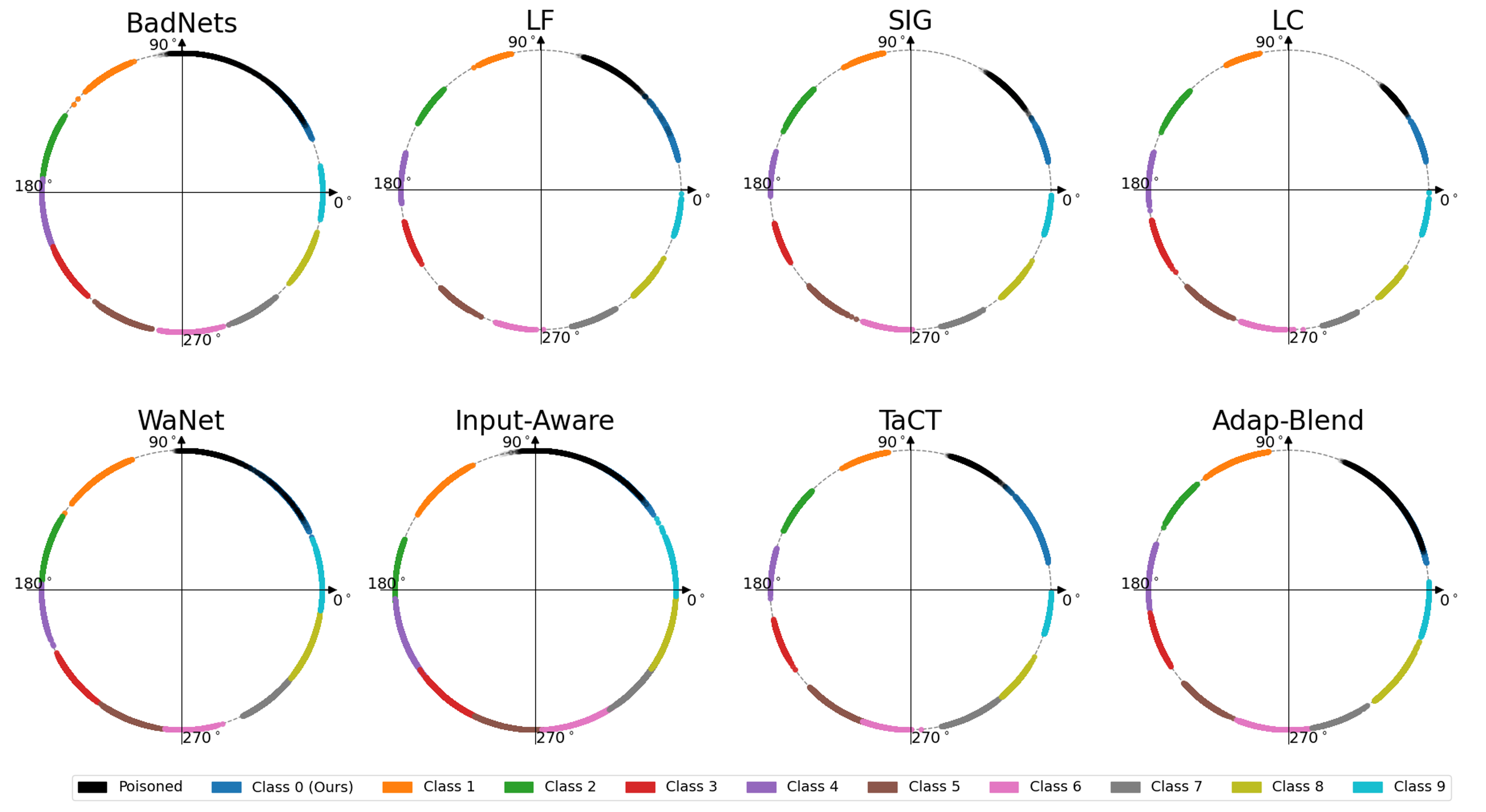}
   \caption{Gradient circular distributions (GCDs) of multiple backdoor attacks on CIFAR-10 with Preact-ResNet18.}
   \label{supple:gcd_preact}
\end{figure}

\begin{figure}[h]
  \centering
	\includegraphics[width=0.8\linewidth]{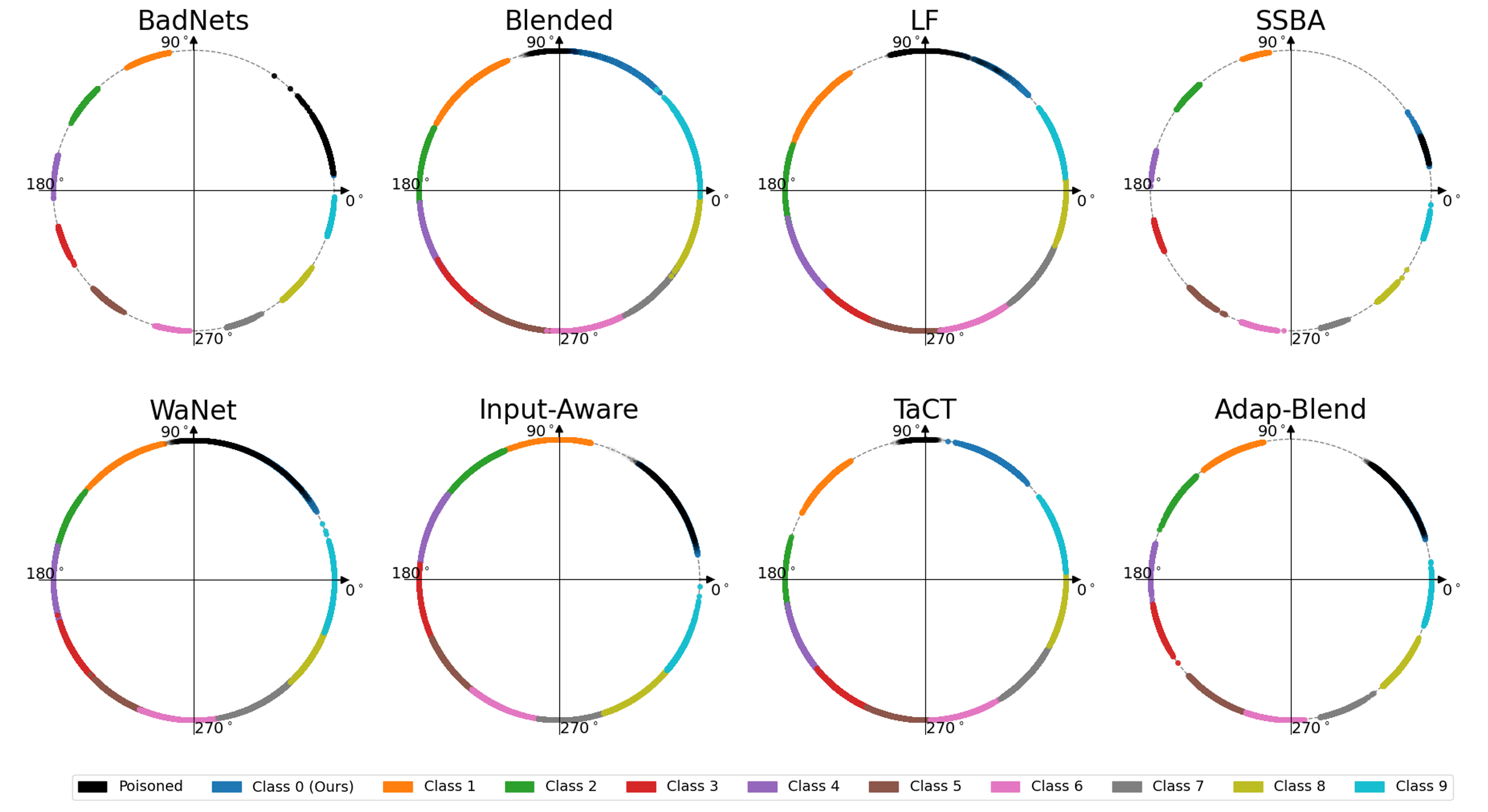}
   \caption{Gradient circular distributions (GCDs) of multiple backdoor attacks on CIFAR-10 with VGG19-BN.}
   \label{supple:gcd_vgg}
\end{figure}

Fig. \ref{supple:gcd_preact} and Fig. \ref{supple:gcd_vgg} present the GCDs for all classes in CIFAR-10 under eight backdoor attacks with 10\% poisoning ratio, based on the model structures of Preact-ResNet18 and VGG19-BN, respectively. It can be noticed that the target class (covering both black and blue arcs) occupies a longer arc on the circle compared with the clean classes across different model structures and backdoor attacks. Note that we moved all clean classes’ arcs to different areas on the circle to avoid visual overlap.

\section{The separation between poisoned and clean samples}
\label{appendix:separation}
\begin{figure}[h]
    \centering
    \includegraphics[width=0.8\linewidth]{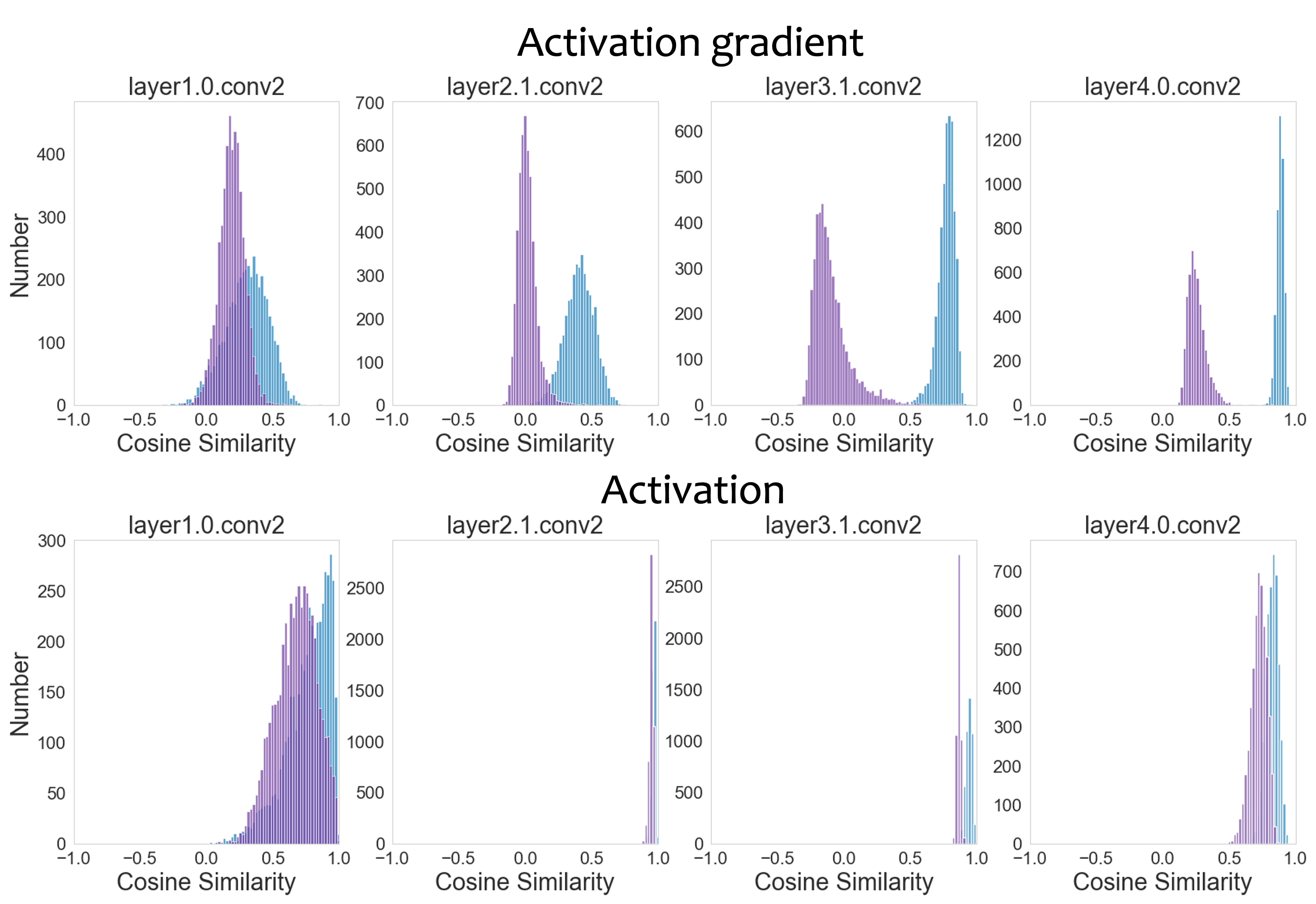}
    \caption{The distribution of the cosine similarities of samples from the target class with a clean sample in multiple convolutional layers. The model structure is Preact-ResNet18. The purple represents poisoned samples, while the blue represents clean ones.}
    \label{fig:seperation}
\end{figure}

Fig. \ref{fig:seperation} displays the cosine similarities of samples in the target class with a clean basis, which are computed in activation gradient and activation spaces, respectively. We exhibit the distributions of cosine similarities of four convolutional layers. If a sample is clean, the cosine similarity should be large. As shown in Fig.~\ref{fig:seperation}, the separation of the distribution of cosine similarities between clean and poisoned samples is larger in the activation gradient space, which can be observed in many convolutional layers. 

\section{The results of the compared ASSET on ResNet18}
In our above experiment, we found that detection performance of ASSET is deviated from the results reported in the paper \cite{pan2023asset}, such as BadNets and Blended attacks. Therefore, we decided to replicate these detection results using the recommended model structure, ResNet18. We tested six different attacks, including BadNets, Blended, WaNet, Input-Aware, TaCT, and Adap-Blend. The experiments were conducted on the CIFAR-10 dataset with 10\% poisoning ratio. The corresponding results are shown in Tab. \ref{tab:asset_resnet183}. When changing the model structure from Preact-ResNet18 to ResNet18, we discovered that the detection performance of ASSET becomes better, such as the TPRs of BadNets, Blended, and TaCT. However, when the form of the trigger is too complex and dynamic, which requires the model to spend more epochs learning it, this poses a greater challenge for the loss-based method, like ABL and ASSET. The work \cite{wu2022backdoorbench} provides more analysis of quick learning of backdoors. 

\label{appendix:asset}
\input{table/appendix/asset_resnet183}

\section{t-SNE results}
\label{appendix:tsne}
\begin{figure}[h]
    \centering
    \includegraphics[width=0.9\linewidth]{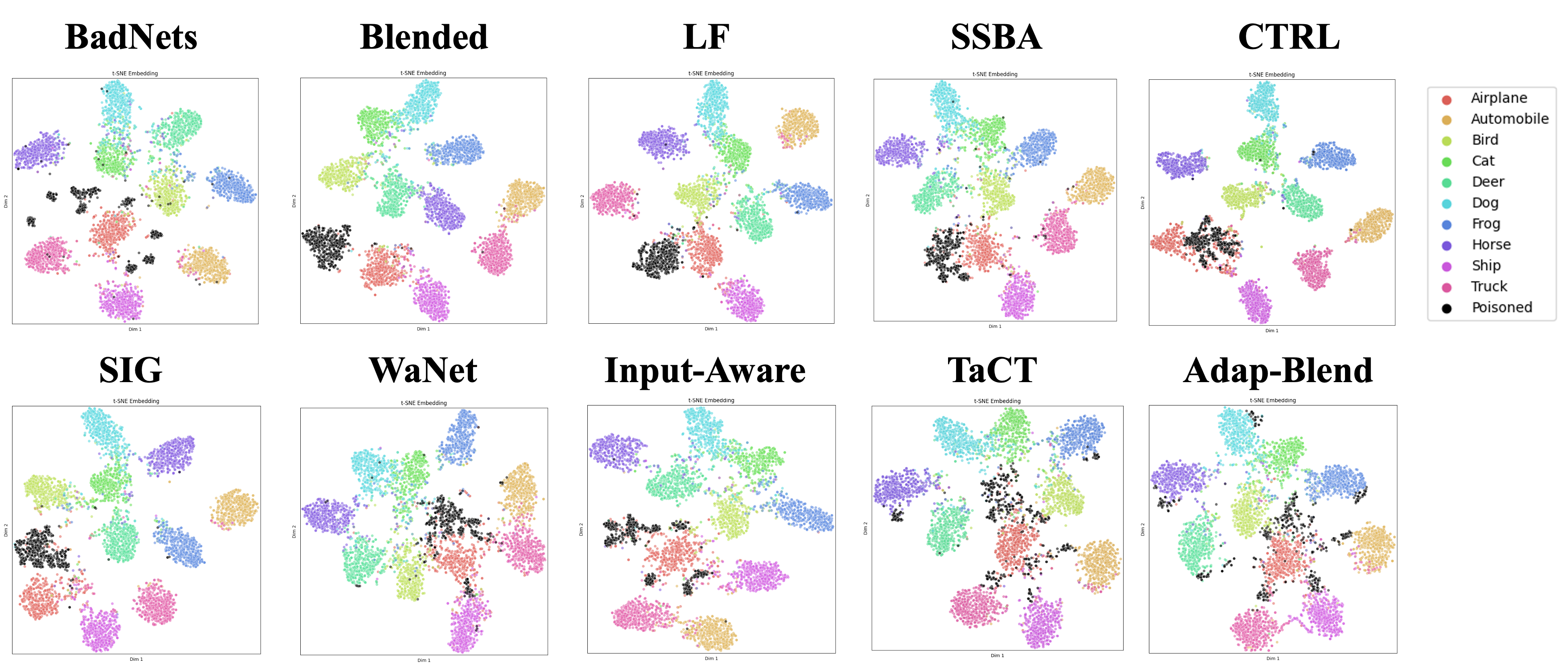}
    \caption{t-SNE visualization of ten backdoor models trained on poisoned CIFAR-10 under 10\% poisoning ratio for non-clean label attacks and 5\% poisoning ratio for clean label attacks. The model architecture is Preact-ResNet18.}
    \label{fig:tsne_preact}
\end{figure}

In this section, we provide the t-SNE visualizations of ten backdoor attacks conducted on CIFAR-10 with Preact-ResNet18 in Fig. \ref{fig:tsne_preact}. The activations are extracted from the last convolutional layer (\textit{layer4.1.conv2}). From Fig \ref{fig:tsne_preact}, it is evident that the dispersion of activations of poisoned and clean samples is significant in some attacks, such as BadNets and Blended. Consequently, most activation-based methods perform well against these attacks. Additionally, the dispersion of activations of the target class is relatively small in CTRL attacks, where these methods fail.

%% file: table/appendix/model_structure.tex
\begin{table}[h]
\centering
\caption{Introduction of PreAct-ResNet18 and VGG19-BN Architectures}
\renewcommand\arraystretch{1.3}
\setlength{\tabcolsep}{4pt}
\scalebox{0.8}{
\begin{tabular}{c|c|c|c}
\toprule
Layer Name & Output Size & PreAct-ResNet18 & VGG19-BN \\
\midrule
Conv1 & 112x112 & 7x7, 64, stride 2 & 3x3, 64 \\
\hline
Max Pooling & 56x56 & 3x3, stride 2 & 2x2, stride 2 \\
\hline
Residual Block 1 & 56x56 & 
\begin{tabular}[c]{@{}c@{}}
\textbf{Pre-activation:} BN, ReLU \\
3x3, 64 \\
\textbf{Pre-activation:} BN, ReLU \\
3x3, 64
\end{tabular} &
\begin{tabular}[c]{@{}c@{}}
3x3, 64 \\
3x3, 64
\end{tabular} \\
\hline
Residual Block 2 & 28x28 & 
\begin{tabular}[c]{@{}c@{}}
\textbf{Pre-activation:} BN, ReLU \\
3x3, 128, stride 2 \\
\textbf{Pre-activation:} BN, ReLU \\
3x3, 128
\end{tabular} &
\begin{tabular}[c]{@{}c@{}}
3x3, 128 \\
3x3, 128
\end{tabular} \\
\hline
Residual Block 3 & 14x14 & 
\begin{tabular}[c]{@{}c@{}}
\textbf{Pre-activation:} BN, ReLU \\
3x3, 256, stride 2 \\
\textbf{Pre-activation:} BN, ReLU \\
3x3, 256
\end{tabular} &
\begin{tabular}[c]{@{}c@{}}
3x3, 256 \\
3x3, 256 \\
3x3, 256 \\
3x3, 256
\end{tabular} \\
\hline
Residual Block 4 & 7x7 & 
\begin{tabular}[c]{@{}c@{}}
\textbf{Pre-activation:} BN, ReLU \\
3x3, 512, stride 2 \\
\textbf{Pre-activation:} BN, ReLU \\
3x3, 512
\end{tabular} &
\begin{tabular}[c]{@{}c@{}}
3x3, 512 \\
3x3, 512 \\
3x3, 512 \\
3x3, 512
\end{tabular} \\
\hline
Global Average Pooling & 1x1 & \multicolumn{2}{c}{Global Average Pooling} \\
\hline
Fully Connected & 1x1 & 1000-d & 4096 \\
\hline
Fully Connected & 1x1 & & 4096 \\
\hline
Fully Connected & 1x1 & & 1000-d \\
\hline
Softmax & 1x1 & \multicolumn{2}{c}{Softmax} \\
\bottomrule
\end{tabular}}
\label{tab:model_structure}
\end{table}

%% file: table/appendix/hyper_attacks.tex
\begin{table}[h]
\centering
\caption{The hyper-parameters of implemented backdoor attacks for CIFAR-10.}
\renewcommand\arraystretch{1.2}
\setlength{\tabcolsep}{3pt} 
\scalebox{0.6}{\begin{tabular}{c|p{3cm}|c|c|p{3cm}|c}
\hline
 & \centering \textbf{Category} & \textbf{Attack} &\textbf{Parameters} & \centering \textbf{Usage} & \textbf{Value}\\
\hline
\multirow{19}{*}{All-to-one} & \multirow{3}{3cm}{\centering non-clean label with sample-agnostic trigger} & BadNets & / & /& /\\
\cline{3-6}
  & & Blended & $\alpha$ & the transparency of the trigger. & 0.2\\
  \cline{3-6}
 & & LF & $\alpha$ & fooling rate & 0.2 \\
\cline{2-6}
  &\multirow{2}{3cm}{\small{\centering clean label with sample-agnostic trigger}} & \multirow{2}{*}{SIG} & $\Delta $ & \multirow{2}{3cm}{to generate sinusoidal signal.}  & 40 \\
    & & &$f$ & & 6 \\
\cline{2-6}
 &\multirow{2}{3cm}{\centering clean label with sample-specific trigger} & CTRL & $c$& trigger channel& [2,1]\\
  &&&$l$&trigger location & (12,27)\\

\cline{2-6}
  & \multirow{5}{3cm}{\centering non-clean label with sample-specific trigger} & SSBA & /& /&/\\
\cline{3-6}
& & \multirow{3}{*}{TaCT} & $s-class$ & the trigger will be added to samples in $s-class$ and change their labels to the target label.  & a list\\
& & & $c-class$ & samples in $c-class$ will only be added the trigger. & a list\\
& & & $c$ & control the number of samples in $c-class$ & 0.1\\
\cline{3-6}
& & Adap-Blend & $m$ & the probability of the area being masked. & 0.5 \\
\cline{2-6}
 & \multirow{5}{3cm}{\centering non-clean label with training control} & \multirow{4}{*}{WaNet} & $s$ & warping strength & 0.5\\
 & & & $k$ & grid scale & 4\\
  & & & $\rho_a$ & backdoor probability & =poisoning ratio\\
   & & & $\rho_n$ & the noise probability & 0.1\\
\cline{3-6}
 & & \multirow{3}{*}{Input-Aware} & $\lambda_{div}$& the diversity enforcement regularisation. & 1\\
  & & &$\rho_b$ &the backdoor probability. & =poisoning ratio\\
   & & & $\rho_c$& the cross-trigger probability. & 0.1 \\
\hline
\multirow{2}{*}{All-to-all}&\centering non-clean label with sample-agnostic trigger&BadNets-A2A& \multirow{2}{*}{$K$}&\multirow{2}{3cm}{to compute target labels.} & \multirow{2}{*}{10}\\
\cline{2-3}
&\centering non-clean label with sample-specific trigger&SSBA-A2A& & & \\
\hline
\multirow{4}{*}{Single-trigger Attack}&\multirow{3}{3cm}{\centering non-clean label with sample-agnostic trigger}&BadNets& \multirow{4}{*}{$C$} & \multirow{4}{3cm}{to compute target labels} & \multirow{4}{*}{5}\\
\cline{3-3}
 & & Blended&  & & \\
 \cline{3-3}
  & & LF&  & & \\
\cline{2-3}
   &\centering non-clean label with sample-specific trigger & SSBA&  &  & \\
\hline
\multirow{3}{*}{Multi-trigger Attack} & \multirow{2}{3cm}{\centering non-clean label with sample-agnostic trigger} & BadNets+Blended+SSBA+ LF+ SIG&\multirow{3}{*}{$r$}  & \small{the ratio of poisoned samples from each type of trigger}  &0.02\\
& & &  & &\\
& & &   & \\
\hline
\end{tabular}
}
\label{tab:hyper_attack}
\end{table}

%% file: table/appendix/complexity_theo.tex
\begin{table}[h]
\centering
\renewcommand\arraystretch{1.3}
\setlength{\tabcolsep}{3pt}
\caption{Computation overhead of AGPD and the compared methods on CIFAR-10. The value in the bracket represent standard deviation.}
\scalebox{0.7}{
\begin{tabular}{c|ccccccccc}
\hline
         & AC         & Beatrix    & SCAn       & Spectral   & STRIP       & ABL         & CD          & ASSET & AGPD       \\ \hline
Time (minute) & 1.02(0.01) & 8.92(0.38) & 1.27(0.02) & 1.73(0.06) & 3.29( 0.02) & 10.06(0.09) & 20.42(0.02) &   206.43 (12.63)    & 5.02(0.03) \\ \hline
\end{tabular}
}
\label{tab:complexity}
\end{table}

%% file: alg/filtering.tex
\begin{algorithm}[h]
	\renewcommand{\algorithmicrequire}{\textbf{Input:}}
	\renewcommand{\algorithmicensure}{\textbf{Output:}}
	\caption{Filtering out poisoned samples within the identified target class(es).} 
	\label{alg:psf} 
	\begin{algorithmic}[1]
        \REQUIRE The identified target class $k^*$, the subset $\mathcal{D}_{bd}^{k^*}$, selected layer $l^*$, the reference $(\x_0,k^*)$, and filtering threshold $\tau_s$.
        \ENSURE Suspected set $\mathcal{D}_{sus}^{k^*}$ and purified set $\mathcal{D}_{bd}^{k^*}\backslash\mathcal{D}_{sus}^{k^*}$.
        \STATE Compute the GCDs of the set $\mathcal{D}_{bd}^{k^*}$, referred to $\{\theta_{\x_0}(\x_i)\}_{i=1}^{n_{k^*}}$, corresponding to the reference $(x_0,k^*)$,  , according to Eq.(\ref{eq: gcd}).
        \STATE Find the farthest activation gradient $g(\x_{n^*})$ according to $n^* = \arg\max_{i\in \{1,\ldots,n_{k^*}\}} \theta_{\x_0}(\x_i)$.
        \STATE Set $\mathcal{D}_{sus}^{k^*} = \emptyset $, $JS = \{\}$, and iteration $t=0$.
        \WHILE{$\mathcal{D}_{bd}^{k^*}\backslash\mathcal{D}_{sus}^{k^*} \neq \emptyset$}
        \STATE Calculate the distribution of $\{s_{\x_0}(\x_i)\}_{\x_i \in \D_{bd}^{k^*}}$ according to Eq.(\ref{eq:trust_score}).
        \STATE Add samples ${(\x_i,k^*)}$ whose $s_{\x_0}(\x_i)$ is smaller than $\tau_s$ to $\mathcal{D}_{sus}^{k^*}$, and remove them from $\mathcal{D}_{bd}^{k^*}$.
        \IF{$t>0$}
        \STATE Calculate the JS divergence between the distribution of $\{s_{\x_0}(\x_i)\}_{\x_i \in \D_{bd}^{k^*}}$ in the iteration $t$ and $t-1$.
        \STATE Add the JS divergence to $JS$.
        \ENDIF
        \STATE $t = t + 1 $.
        \ENDWHILE
        \STATE Find an appropriate iteration $t^*$ according to the stopping criteria.
        \STATE Remain samples ${(\x_i,y)}$ filtered out before the iteration $t^*$ in $\mathcal{D}_{sus}^{k^*}$. 

	\end{algorithmic} 
\end{algorithm}

%% file: table/appendix/dtd_gtsrb_result_fscore.tex
\begin{table}[h]
\centering
\caption{The detection performance of AGPD and compared detectors on DTD and GTSRB. The results are evaluated on Preact-ResNet18.}
\renewcommand\arraystretch{1.3}
\setlength{\tabcolsep}{3pt} 
\scalebox{0.5}{
\begin{tabular}{c|c|c|ccc|ccc|ccc|ccc|ccc}
\toprule
\multirow{2}{*}{Dataset}       & \multirow{2}{*}{Attack} & Backdoored  & \multicolumn{3}{c|}{SCAn~\cite{tang2021demon}} & \multicolumn{3}{c|}{STRIP~\cite{gao2019strip}} & \multicolumn{3}{c|}{ABL~\cite{li2021anti}} & \multicolumn{3}{c|}{ASSET~\cite{pan2023asset}} & \multicolumn{2}{c}{AGPD} \\

&                         & ACC/ASR     & TPR$\uparrow$     & FPR$\downarrow$    & F1$\uparrow$ & TPR$\uparrow$     & FPR$\downarrow$      & F1$\uparrow$ & TPR$\uparrow$      & FPR$\downarrow$    & F1$\uparrow$ & TPR$\uparrow$     & FPR$\downarrow$     & F1$\uparrow$ & TPR$\uparrow$     & FPR$\downarrow$     & F1$\uparrow$\\
\hline
\multirow{5}{*}{\rotatebox{90}{DTD}}

&BadNets \cite{gu2019badnets}&51.97/98.32&90.43&\textbf{0.21}&\textbf{94.05}&83.78&12.71&56.20&76.33&2.63&76.43&\underline{96.01}&8.16&71.15&\textbf{99.47}&\underline{1.60}&\underline{93.03}\\
&Blended \cite{chen2017targeted}&51.86/94.62&82.18&\textbf{0.30}&88.76&\underline{95.74}&13.21&60.74&88.83&\underline{1.24}&\underline{88.95}&0.80&3.04&1.25&\textbf{100.0}&1.89&\textbf{92.27}\\
&WaNet \cite{nguyen2021wanet}&42.71/26.41&\underline{86.08}&\textbf{1.06}&\underline{88.01}&77.84&14.03&51.14&0.28&11.00&0.27&1.70&2.33&2.61&\textbf{100.0}&\underline{1.88}&\textbf{92.27}\\
&\small{Input-Aware~\cite{nguyen2020input}}&45.85/85.54&0.00&\textbf{0.00}&0.00&13.07&12.91&11.38&\underline{20.45}&8.92&\underline{20.19}&0.00&0.72&0.00&\textbf{98.58}&\underline{0.59}&\textbf{96.73}\\
&\small{Adap-Blend \cite{adapt}}&49.41/85.65&\underline{77.13}&\textbf{0.92}&\underline{83.21}&32.18&16.19&23.01&6.91&10.34&6.67&4.26&2.99&6.49&\textbf{100.0}&\underline{1.18}&\textbf{95.07}\\
\rowcolor{gray!20}
&Avg.&&\underline{67.16}&\textbf{0.50}&\underline{70.81}&60.52&13.81&40.50&38.56&6.83&38.50&20.55&3.45&16.30&\textbf{99.61}&\underline{1.43}&\textbf{93.88}\\

\midrule
\midrule
\multirow{5}{*}{\rotatebox{90}{GTSRB}}&BadNets \cite{gu2019badnets}&96.35/95.02&94.62&4.00&\underline{82.06}&\underline{95.54}&15.39&57.20&73.71&\underline{2.92}&73.71&\textbf{100.0}&47.79&31.74&\textbf{100.0}&\textbf{0.27}&\textbf{98.80}\\
&Blended \cite{chen2017targeted}&98.17/100.0&83.47&7.54&66.42&\textbf{100.0}&11.58&65.74&80.54&\underline{2.16}&\underline{80.55}&\underline{99.36}&41.33&34.77&95.87&\textbf{0.30}&\textbf{96.57}\\
&WaNet \cite{nguyen2021wanet}&97.05/96.16&\underline{60.83}&\textbf{0.00}&\underline{75.63}&7.83&14.34&6.59&0.00&11.03&0.00&44.07&64.79&12.12&\textbf{100.0}&\underline{0.20}&\textbf{99.12}\\
&\small{Input-Aware~\cite{nguyen2020input}}&97.91/95.64&\underline{52.15}&\textbf{0.00}&\underline{68.54}&1.99&\underline{10.44}&2.03&0.00&11.03&0.00&3.07&59.18&0.96&\textbf{79.60}&\textbf{0.00}&\textbf{88.64}\\
&\small{Adap-Blend \cite{adapt}}&97.66/80.42&50.74&\underline{3.95}&\underline{54.48}&20.48&12.40&17.63&0.00&11.11&0.00&\underline{81.10}&22.48&42.30&\textbf{93.44}&\textbf{0.23}&\textbf{95.58}\\
\rowcolor{gray!20}
&Avg.&&\underline{68.36}&\underline{3.10}&\underline{69.43}&45.17&12.83&29.84&30.85&7.65&30.85&65.52&47.12&24.38&\textbf{93.78}&\textbf{0.20}&\textbf{95.74}\\

\bottomrule

\end{tabular}
}
\label{tab:other_dataset}
\end{table}

%% file: table/appendix/imagenetsub.tex
\begin{table}[h]
\centering
\caption{The detection performance of AGPD on ImageNet-200. The results are evaluated on Preact-ResNet18.}
\scalebox{0.8}{
\begin{tabular}{c|c|c|ccc}
\toprule
\multirow{2}{*}{Dataset}      & \multirow{2}{*}{Attack} & Backdoored  & \multicolumn{3}{c}{AGPD}     \\
                              &                         & ACC/ASR     & TPR$\uparrow$ & FPR $\downarrow$ & F1 $\uparrow$ \\ \midrule
\multirow{4}{*}{ImageNet-200} & BadNet                  & 78.57/80.03 & 94.62  & 0.00 & 97.24  \\
                              & Blended                 & 79.95/99.93 & 100.0 & 0.47 & 97.93  \\
                              & Adap-Blend              & 72.3/93.17  & 99.98  & 0.59 & 99.19  \\
\rowcolor{gray!20}
&Avg. & & 		98.20&	0.35 &	98.12\\

                              \bottomrule
\end{tabular}}
\label{tab:imagenet}
\end{table}

%% file: table/appendix/agpd_vgg19_fscore.tex
\begin{table}[h]
\centering
\caption{The detection performance of AGPD and compared detectors on CIFAR-10 and Tiny ImageNet. The results are evaluated on VGG19-BN. }
\renewcommand\arraystretch{1.3}
\setlength{\tabcolsep}{2pt} 
\scalebox{0.45}{
\begin{tabular}{c|c|c|ccc|ccc|ccc|ccc|ccc|ccc|ccc|ccc|ccc}
\toprule
\multirow{2}{*}{Dataset}       & \multirow{2}{*}{Attack} & Backdoored  & \multicolumn{3}{c|}{AC~\cite{ac}} & \multicolumn{3}{c|}{Beatrix~\cite{mabeatrix}} & \multicolumn{3}{c|}{SCAn~\cite{tang2021demon}} & \multicolumn{3}{c|}{Spectral~\cite{tran2018spectral}} & \multicolumn{3}{c|}{STRIP~\cite{gao2019strip}} & \multicolumn{3}{c|}{ABL~\cite{li2021anti}} & \multicolumn{3}{c|}{CD~\cite{huang2023distilling}} & \multicolumn{3}{c|}{ASSET~\cite{pan2023asset}} & \multicolumn{2}{c}{AGPD} \\

&                         & ACC/ASR     & TPR$\uparrow$     & FPR$\downarrow$    & F1$\uparrow$ & TPR$\uparrow$     & FPR$\downarrow$      & F1$\uparrow$ & TPR$\uparrow$      & FPR$\downarrow$    & F1$\uparrow$ & TPR$\uparrow$     & FPR$\downarrow$     & F1$\uparrow$ & TPR$\uparrow$      & FPR$\downarrow$     & F1$\uparrow$ & TPR$\uparrow$     & FPR$\downarrow$     & F1$\uparrow$ & TPR$\uparrow$     & FPR$\downarrow$     & F1$\uparrow$ & TPR$\uparrow$     & FPR$\downarrow$     & F1$\uparrow$ & TPR$\uparrow$      & FPR$\downarrow$    & F1$\uparrow$ \\
\hline
\multirow{13}{*}{\rotatebox{90}{CIFAR-10}}&BadNets \cite{gu2019badnets}&91.82/93.79&0.00&6.42&0.00&52.90&3.44&57.55&68.90&\underline{0.35}&\textbf{80.08}&28.28&\textbf{0.02}&44.02&\underline{82.88}&10.84&59.10&76.68&2.59&\underline{76.68}&72.98&13.46&49.62&4.54&2.33&7.24&\textbf{99.72}&6.90&76.16\\
&Blended \cite{chen2017targeted}&93.69/99.75&0.00&14.79&0.00&99.22&9.97&68.68&96.72&\textbf{0.00}&\underline{98.33}&28.50&\textbf{0.00}&44.36&47.06&10.74&38.61&78.08&2.44&78.08&\underline{99.94}&99.87&18.19&0.20&52.35&0.07&\textbf{99.96}&\underline{0.02}&\textbf{99.90}\\
&LF \cite{zeng2021rethinking}&93.01/99.05&0.00&6.76&0.00&\underline{99.46}&6.40&77.39&83.22&\textbf{0.00}&\underline{90.83}&28.50&\textbf{0.00}&44.36&89.80&8.28&67.94&40.80&6.58&40.80&0.10&\underline{0.03}&0.20&1.74&0.97&3.15&\textbf{99.50}&\underline{0.03}&\textbf{99.62}\\
&SSBA \cite{li2021invisible}&92.88/97.06&4.74&14.08&4.10&0.10&6.19&0.13&\underline{89.76}&\underline{0.80}&\textbf{91.15}&2.18&2.92&3.39&72.38&12.33&51.08&25.16&8.32&25.16&84.78&60.67&23.20&0.00&\textbf{0.00}&0.00&\textbf{98.94}&4.81&\underline{81.69}\\
&SIG \cite{SIG}&93.40/95.43&0.00&13.84&0.00&\textbf{100.0}&3.56&74.70&94.80&\textbf{0.00}&\underline{97.33}&30.00&\textbf{0.00}&46.15&98.96&6.20&62.47&94.84&5.53&63.23&88.24&48.16&28.39&\textbf{100.0}&47.35&18.19&\underline{99.60}&\underline{0.02}&\textbf{99.62}\\
&CTRL&95.52/98.8&15.52&17.08&7.06&22.20&4.79&20.83&5.80&\textbf{0.00}&10.96&23.20&14.57&11.60&\underline{94.36}&10.19&48.64&86.72&5.96&57.81&88.16&23.29&44.33&\textbf{100.0}&5.81&\underline{64.42}&90.84&\underline{2.06}&\textbf{78.99}\\
&WaNet \cite{nguyen2021wanet}&89.68/96.94&0.00&20.48&0.00&2.45&9.64&2.51&0.00&\textbf{0.00}&0.00&29.03&\underline{0.06}&\underline{44.81}&1.92&12.02&1.76&13.18&9.67&12.76&\textbf{98.85}&98.54&18.21&0.00&38.57&0.00&\underline{85.35}&0.09&\textbf{91.67}\\
&\small{Input-Aware~\cite{nguyen2020input}}&90.82/98.17&10.52&7.01&11.80&2.03&5.63&2.59&65.04&\textbf{0.03}&\underline{78.69}&1.88&2.86&2.90&0.98&8.60&1.07&14.83&9.50&14.35&\underline{84.26}&58.53&23.70&0.06&10.01&0.07&\textbf{92.96}&\underline{0.06}&\textbf{96.09}\\
&TaCT \cite{tang2021demon}&93.21/95.95&0.00&6.46&0.00&97.24&6.79&75.29&\underline{99.22}&\textbf{0.00}&\textbf{99.61}&30.00&\textbf{0.00}&46.15&75.30&14.19&49.70&26.94&8.12&26.94&83.42&74.00&19.64&0.42&0.66&0.79&\textbf{100.0}&\underline{0.25}&\underline{98.90}\\
&\small{Adap-Blend \cite{adapt}}&92.87/66.17&4.68&16.07&3.75&0.22&7.21&0.27&29.68&\underline{1.80}&\underline{40.68}&1.36&3.03&2.12&7.34&11.31&7.02&0.02&11.11&0.02&\textbf{100.0}&100.0&18.18&0.00&\textbf{0.01}&0.00&\underline{99.92}&8.17&\textbf{73.07}\\
&BadNets-A2A&91.93/74.40&\underline{79.46}&\underline{0.02}&\underline{88.49}&28.42&6.66&30.17&0.00&\textbf{0.00}&0.00&0.00&1.67&0.00&1.54&11.72&1.49&4.86&10.57&4.86&58.16&20.01&34.39&5.86&3.67&8.44&\textbf{95.92}&\underline{0.02}&\textbf{97.83}\\
&SSBA-A2A&93.46/87.84&\underline{99.36}&0.93&\underline{95.66}&31.44&7.17&32.08&0.00&\textbf{0.00}&0.00&0.00&1.67&0.00&19.98&16.14&15.07&1.18&10.98&1.18&\textbf{99.98}&99.96&18.18&0.24&0.11&0.47&94.82&\underline{0.02}&\textbf{97.27}\\
\rowcolor{gray!20}
&Avg.&&17.86&10.33&17.57&44.64&6.45&36.85&52.76&\textbf{0.25}&\underline{57.31}&16.91&2.23&24.15&49.38&11.05&33.66&38.61&7.61&33.49&\underline{79.91}&58.04&24.69&17.76&13.49&8.57&\textbf{96.46}&\underline{1.87}&\textbf{90.90}\\

\midrule
\midrule
\multirow{9}{*}{\rotatebox{90}{Tiny ImageNet}}&BadNets \cite{gu2019badnets}&56.12/99.90&0.10&20.08&0.07&97.73&9.37&69.30&99.88&\textbf{0.00}&\textbf{99.94}&15.67&\textbf{0.00}&27.09&\underline{99.99}&11.24&66.41&96.66&0.37&96.66&96.43&9.16&69.15&\textbf{100.0}&48.50&31.42&99.67&\underline{0.12}&\underline{99.30}\\
&Blended \cite{chen2017targeted}&55.53/97.57&0.00&7.47&0.00&92.85&4.40&79.90&78.26&\textbf{0.00}&87.80&15.67&\textbf{0.00}&27.09&95.55&13.88&59.63&\underline{96.78}&0.36&\underline{96.78}&82.38&34.19&33.62&1.32&4.65&1.84&\textbf{99.99}&\underline{0.03}&\textbf{99.85}\\
&LF \cite{zeng2021rethinking}&55.21/98.51&0.00&5.74&0.00&22.73&2.02&32.26&52.25&\textbf{0.00}&\underline{68.64}&15.67&\textbf{0.00}&27.09&21.09&11.82&18.55&48.83&5.69&48.83&96.43&10.55&66.18&\underline{99.08}&17.03&56.24&\textbf{99.92}&\underline{0.03}&\textbf{99.84}\\
&SSBA \cite{li2021invisible}&55.97/97.69&0.00&8.88&0.00&42.53&3.96&47.73&78.05&\textbf{0.00}&87.67&15.67&\textbf{0.00}&27.09&99.92&11.77&65.33&89.65&1.15&\underline{89.65}&93.11&19.00&51.14&\textbf{100.0}&47.66&31.80&\underline{99.94}&\underline{0.02}&\textbf{99.86}\\
&WaNet \cite{nguyen2021wanet}&58.33/90.35&0.02&5.62&0.03&0.00&1.51&0.00&99.94&\textbf{0.00}&\textbf{99.96}&13.93&0.19&24.08&17.65&11.57&15.38&83.54&2.39&80.85&99.09&0.30&98.19&\textbf{100.0}&0.73&97.28&\underline{99.97}&\underline{0.12}&\underline{99.40}\\
&\small{Input-Aware~\cite{nguyen2020input}}&57.5/99.75&0.05&5.64&0.07&0.53&\underline{2.90}&0.83&\textbf{98.11}&\textbf{0.00}&\textbf{99.05}&15.72&\textbf{0.00}&27.17&15.49&12.35&13.19&64.97&4.31&62.87&95.88&8.36&70.73&\underline{97.18}&15.79&58.19&82.54&\textbf{0.00}&\underline{90.43}\\
&TaCT \cite{tang2021demon}&54.93/91.25&0.00&8.16&0.00&54.87&2.62&61.50&44.87&\textbf{0.00}&\underline{61.95}&15.05&\underline{0.08}&26.00&\underline{99.91}&17.27&56.24&56.75&4.81&56.75&74.04&42.16&26.75&99.35&51.83&31.05&\textbf{99.99}&0.24&\textbf{98.94}\\
&\small{Adap-Blend \cite{adapt}}&54.55/96.35&25.95&21.18&16.39&0.68&10.67&0.69&34.92&\textbf{0.01}&\underline{51.73}&14.93&\underline{0.08}&25.81&67.20&15.69&43.58&1.09&10.99&1.09&81.50&23.87&41.12&\textbf{100.0}&57.36&30.35&\underline{99.99}&0.16&\textbf{99.27}\\
\rowcolor{gray!20}
&Avg.&&3.27&10.35&2.07&38.99&4.68&36.53&73.28&\textbf{0.00}&\underline{82.09}&15.29&\underline{0.04}&26.43&64.60&13.20&42.29&67.28&3.76&66.69&\underline{89.86}&18.45&57.11&87.12&30.44&42.27&\textbf{97.75}&0.09&\textbf{98.36}\\

\bottomrule

\end{tabular}
}
\label{tab:vgg_main}
\end{table}

%% file: table/appendix/asset_resnet183.tex
\begin{table}[]
\centering
\caption{The results of ASSET on CIFAR-10 with ResNet18.}
\renewcommand\arraystretch{1.3}
\setlength{\tabcolsep}{3pt}
\scalebox{0.8}{
\begin{tabular}{ccccccc}
\hline
    & BadNets & Blended & WaNet  & Input-Aware & TaCT     & Adap-Blend \\ \hline
TPR & 90.70 & 99.90 & 1.88 & 0.25      & 100.00 & 54.58    \\
FPR & 0.18  & 9.43  & 1.69 & 0.22      & 0.00   & 38.90    \\
F1  & 94.32 & 69.97 & 3.55 & 0.48      & 100.00 & 23.43    \\ \hline
\end{tabular}
}
\label{tab:asset_resnet183}
\end{table}